\RequirePackage{fix-cm}
\documentclass{svjour3}                     % onecolumn (standard format)
\smartqed  % flush right qed marks, e.g. at end of proof
\usepackage{xcolor}
\usepackage{graphicx}
\usepackage{hyperref}
\usepackage{threeparttable}
\usepackage{ amssymb }
\newcommand{\teff}{$T_{\mathrm{eff}}$}
\newcommand{\logg}{\mbox{log \textit{g}}}
\begin{document}

\title{Beryllium abundances in turn-off stars of globular clusters with the CUBES spectrograph\thanks{The authors acknowledge support by the National Science Centre, Poland, through project 2018/31/B/ST9/01469}%\thanks{Grants or other notes
}
%\subtitle{Beryllium in globular clusters}


\author{Riano E. Giribaldi         \and
        Rodolfo Smiljanic %etc.
}

\authorrunning{Giribaldi \& Smiljanic} % if too long for running head

\institute{R.~E.~Giribaldi \& R.~Smiljanic \at
              Nicolaus Copernicus Astronomical Center, Polish Academy of Sciences, ul.~Bartycka 18, 00-716, Warsaw, Poland\\
%              Tel.: +123-45-678910\\
%              Fax: +123-45-678910\\
              \email{riano@camk.edu.pl}           %  \\
}

\date{Received: 21 September 2021/ Accepted 15 March 2022 in Experimental Astronomy.}

\maketitle

%So far, none of the proposed scenarios can explain all observational properties of globular clusters. 
\begin{abstract}
{Globular clusters host multiple stellar populations that display star-to-star variation of light elements that are affected by hot hydrogen burning (e.g., He, C, N, O). Several scenarios have been suggested to explain these variations. Most involve multiple star formation episodes, where later generations are born from material contaminated by the nucleosynthetic products of the previous stellar generation(s). One difficulty in the modelling of such scenarios is knowing the extent to which processed and pristine material are mixed. In this context, beryllium abundances measured in turn-off stars of different generations can provide new information. Beryllium originates from cosmic-ray spallation and can only be destroyed inside stars. Beryllium abundances can thus directly measure the degree of pollution of the material that formed stars in globular clusters. Turn-off stars in globular clusters are however faint and such studies are beyond the capabilities of current instrumentation. In this work, we show the progress that the CUBES spectrograph will bring to this area. Our simulations indicate that CUBES will enable the detection of variations of about 0.6 dex in the Be abundances between stars from different generations, in several nearby globular clusters with turn-off magnitude down to $V$ = 18 mag.  
}
\end{abstract}

\section{Introduction}
\label{intro}
Globular clusters are considered key objects for studying the evolution of the primitive Galaxy. They offer the opportunity of tracing the early Galactic chemical enrichment at well determined ages.
However, globular clusters are complex systems that host multiple stellar populations. This has been shown both with spectroscopy, with the detection of star-to-star abundance variation \cite{1978ApJ...223..487C,2018ARA&A..56...83B}, and with photometry, with the detection of separated sequences in the subgiant and red giant branches in the color magnitude diagram of several clusters \cite{1999Natur.402...55L,2020A&ARv..28....5C}. 
Such signatures have been attributed to multiple episodes of star formation. Consecutive stellar generations have been contaminated by material processed in hot hydrogen burning taking place in the previous generations. This pollution results in the enhancement of He \cite{bedin2004ApJ...605L.125B,piotto2007ApJ...661L..53P} as well as in variation of elements like Li, C, N, O, and Na \cite{kraft1994PASP..106..553K,gratton2007A&A...464..953G,carretta2015ApJ...810..148C,pancino2017A&A...601A.112P}. In particular cases, even heavier elements (Mg, Al, Si, K, Ca, and Sc) have been shown to be affected \cite{2012ApJ...760...86C,2020MNRAS.492.1641M,2021A&A...646A...9C}.

Several scenarios have been proposed to explain the origin of the polluting material affecting the abundances of globular cluster stars. Explosive nucleosynthesis from massive stars is usually discarded, as the material would probably show variation in the abundance of Fe peak elements and it seems difficult for the proto cluster to retain the high velocity ejecta. More likely sources of the polluting material, which have been discussed in the literature, include asymptotic giant branch (AGB) stars \cite{1981ApJ...245L..79C,2009A&A...499..835V}, fast rotating massive stars \cite{2007A&A...464.1029D,2013A&A...552A.121K}, and massive binaries \cite{2009A&A...507L...1D}. A few other alternative scenarios have also been proposed \cite{2018ARA&A..56...83B}. However, it seems that so far all of the proposed scenarios face problems to reproduce key observational facts regarding globular clusters \cite{2018ARA&A..56...83B,2019A&ARv..27....8G,2020A&ARv..28....5C}.

One difficulty in the study of the formation of globular clusters is to understand to which extent processed and pristine material are mixed to form the new stellar generations. Lithium abundances have been used as a critical diagnostics of such mixing \cite{2002A&A...390...91B,2010A&A...524L...2S}. Lithium is a fragile element quickly destroyed at temperatures above $\sim$ 2.5$\times$10$^6$ K. Thus, material processed by hot hydrogen burning is expected to be free of Li. First generation stars have Li abundances compatible with field stars. Subsequent generations are depleted in Li to an extent that reflects the fractions of processed and pristine material that has been mixed before the new episodes of star formation \cite{2019A&ARv..27....8G}. Nevertheless, at least AGB stars, one of the proposed sources of the processed material, can produce and eject new Li \cite{2010MNRAS.402L..72V}, complicating the interpretation of the Li observations \cite{2014ApJ...791...39D}.

In this regard, the measurement of beryllium abundances in turn-off stars of globular clusters might offer an unambiguous alternative to estimate the pollution-dilution of material forming new stellar generations. Beryllium is a light element that can only be produced by cosmic-ray spallation in the interstellar medium (ISM) \cite{2012A&A...542A..67P}. In stars, Be is only destroyed, by proton-capture reactions at temperatures above $\sim$ 3.5$\times$10$^6$ K. The material ejected by any type of evolved star is thus expected to be Be free. Variations in the abundance of Be in turn-off stars, which are at a stage before the convective envelope deepens, are then expected to directly reflect the fraction of polluted material used in the formation of the multiple stellar populations. This conclusion considers that no significant Be production took place in the stellar ejecta that pollutes second generation stars. Spallation in this material can happen only due to the Galactic cosmic-ray flux, as in principle there is no local source able to accelerate particles to cosmic-ray-like energies. The ejecta would be exposed to cosmic-rays for a time scale that is too short to produce Be abundance at the level found in the ISM, particularly given that second generation stars have about the same C+N+O abundances of first generation stars \cite{2020A&ARv..28....5C}.

However, with current instrumentation, Be abundances can not be easily determined in turn-off stars of globular clusters. The Be lines useful for abundance analysis are either in the ground-accessible near ultraviolet (near UV) region ($\sim$ 3130 \AA) or at the UV that is only accessible from space ($\sim$ 2348 \AA). Turn-off stars in globular clusters are relatively faint ($V$ $\geq$ 16-17 mag). Obtaining high-resolution, high signal-to-noise (SNR) spectra in the region of the Be lines requires prohibitively long exposures.

Attempts to detect the Be lines at $\sim$ 3130\AA\ in turn-off stars of the globular clusters NGC~6397 and NGC~6352 have been reported in \cite{pasquini2004A&A...426..651P,pasquini2007A&A...464..601P,2014A&A...563A...3P}. Such observations were performed with UVES, the UV-visual echelle spectrograph \cite{2000SPIE.4008..534D}, at the 8m Very Large Telescope (VLT) of the European Southern Observatory (ESO). A total of 8 exposures of 90 min resulted in SNR $\sim$ 8-15 around the Be lines. Interestingly, all the stars studied presented Be abundances that are consistent with field stars of the same metallicity. This was true even for one of the stars that presents a significantly depleted oxygen abundance, a signature of second generation stars, NGC~6397 --A228. Nevertheless, the abundance uncertainties are high and an investigation of a larger sample is needed to ascertain if there are star-to-star variations of the Be abundances.

It is expected that this type of investigation will become possible with the forthcoming Cassegrain U-Band Efficient Spectrograph (CUBES) \cite{Zanutta_CUBES}. CUBES is a new spectrograph designed to have high throughput in the near-UV (3000-4000 \AA). It will obtain low- (R $\sim$ 6\,000) and medium-resolution (R $\sim$ 24\,000) spectra with the use of two interchangeable image slicers \cite{Calcines_CUBES}. CUBES will be installed at the VLT and is expected to be competitive against ESO's red-optimized 39m extremely large telescope (ELT) at the near-UV region \cite{2014Ap&SS.354..121P,2016SPIE.9908E..9JE}. The science cases motivating an instrument like CUBES have been recently summarized in \cite{2018SPIE10702E..2EE,2020SPIE11447E..60E} and are discussed in detail in several contributions in this Special Issue. 

The goal of this paper is to demonstrate, with spectral simulations, that CUBES will allow the investigation of Be abundances in turn-off stars of the brightest nearby globular clusters with feasible integration times. This is a science case that can not be addressed with current instrumentation and where CUBES will make an unique contribution. The paper is divided as follows. In Section \ref{sec:data}, we describe the data and codes used in the simulations. In Section \ref{sec:science}, we discuss the expected results that can be obtained from CUBES observations of stars in several nearby globular clusters. Finally, Section \ref{sec:summary} summarizes our investigation.

\section{Data}
\label{sec:data}

Synthetic spectra were produced with the radiative transfer code Turbospectrum \cite{turbospectrum} version v19.1\footnote{Available at \url{https://www.lupm.in2p3.fr/users/plez/}} using one-dimensional (1D) MARCS model atmospheres computed in the local thermodynamical equilibrium (LTE) approximation \cite{gustafson2008}. Our line list was built with data of atomic lines extracted from the VALD~3 database \cite{Ryabchikova2015PhyS...90e4005R}. We used the ``Extract Stellar" format, considering solar effective temperature (\teff) and surface gravity (\logg) values plus a detection threshold\footnote{Depth of the weakest line with respect to the deepness of the strongest line produced} of 0.01. All molecular lines in this selection were removed after the extraction. For the molecular lines, we considered instead the following species with data from the mentioned references\footnote{Molecular line lists from these references already formatted for use with Turbospectrum can be downloaded at \url{https://nextcloud.lupm.in2p3.fr/s/r8pXijD39YLzw5T}}:
NH \cite{kurucz1992RMxAA..23...45K}, C$^{12}$H \cite{masseron2014A&A...571A..47M}, C$^{12}$N$^{14}$ \cite{brooke2014ApJS..210...23B}, C$^{13}$N$^{14}$ \cite{sneden2014ApJS..214...26S}, C$^{12}$2 \cite{brooke2013JQSRT.124...11B}, OH \cite{kurucz1992RMxAA..23...45K}, and SiH \cite{kurucz1992RMxAA..23...45K}.

The line list was further processed to remove what we call ``phantom lines", as is described in Section \ref{sec:linelist} below. For completeness, we quickly mention that the parameters of the Ge I line at $\lambda3039.067$~\AA\ were updated with the more accurate values from \cite{peterson2020A&A...638A..64P}. This same line list was used in the simulations discussed in the companion paper by Smiljanic et al. (in this Special Issue).

To simulate the SNR that can be obtained with the CUBES observations, we relied on the CUBES end-to-end simulator (E2E)\footnote{\url{https://cubes.inaf.it/end-to-end-simulator}} \cite{genoni_CUBES}. The E2E helps to understand the expected properties of the observed spectra, by simulating the effects of the atmosphere, the fore-optics, the image slicers, the spectrograph, and the detector. We assumed observations with 3000s of exposure and 600s of overhead, adding up to an observing block of 1 hour (as is the standard in service mode observations with the VLT). Default values for the air mass, precipitable water vapour, and days from new moon were assumed (i.e. 1.16, 2.5, and 0, respectively). For use with the E2E, spectra were computed with 1 \AA\ sampling and a large enough wavelength range for the E2E to estimate the magnitude in the $V$ band. According to the E2E simulations, CUBES spectra will have $R = \lambda/\Delta\lambda = 23\,000$ and a sampling of $\sim$2.35 pixels (pixel size of 0.058~\AA) around the Be lines.

\subsection{Refinement of the line list}
\label{sec:linelist}

Our knowledge of the spectral lines in the near UV is still quite incomplete; several lines remain unidentified. The opposite case is also observed, although in much lower extent, i.e., synthetic spectra can show lines that are not present in observed spectra (called ``phantom lines" henceforth). Such lines can be an issue for precision spectroscopy, as they will cause misinterpretation of the strength of the observed lines. Phantom lines can also affect photometric magnitudes computed from synthetic flux distributions due to an additional decrease in the flux budget \cite{bell1994MNRAS.268..771B}.

We refined the line list by clipping phantom lines identified from a direct comparison of the synthetic and the observed solar spectra. For this purpose, we produced solar synthetic spectra of atomic lines and of each one of the molecular species separately, for an individual analysis. For the solar parameters, we adopted the values that best fit the observed spectra with 1D LTE model atmospheres in the analysis presented by \cite{jofre2014A&A...564A.133J}: \teff~= 5771~K, \logg\ = 4.44~dex, [Fe/H] = 0.03~dex, and microturbulent velocity $v_{mic} = 0.85~\mathrm{ms}^{-1}$.

\begin{figure}
    \centering
    \includegraphics[width=7.5cm]{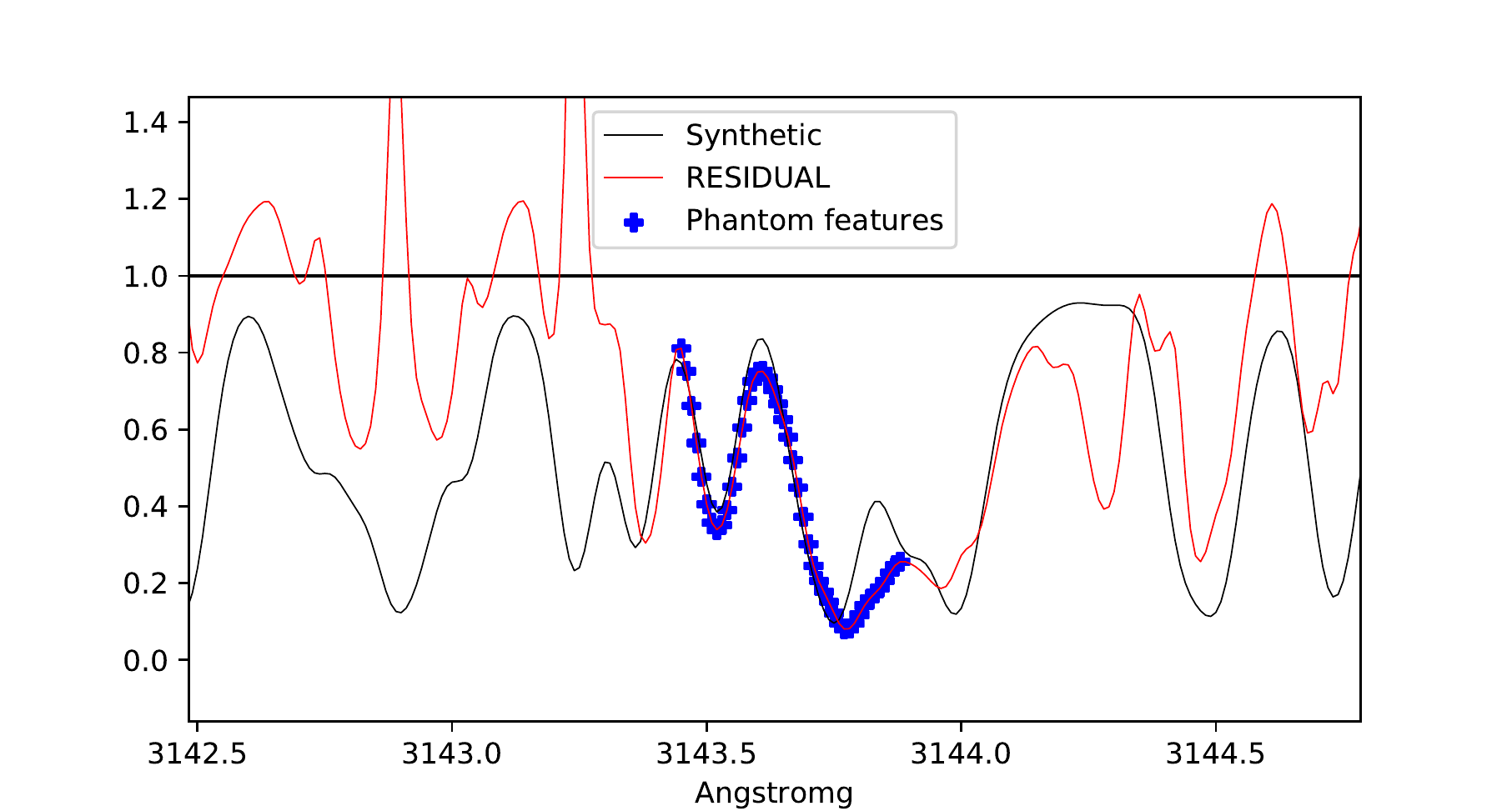}
    \includegraphics[width=8cm]{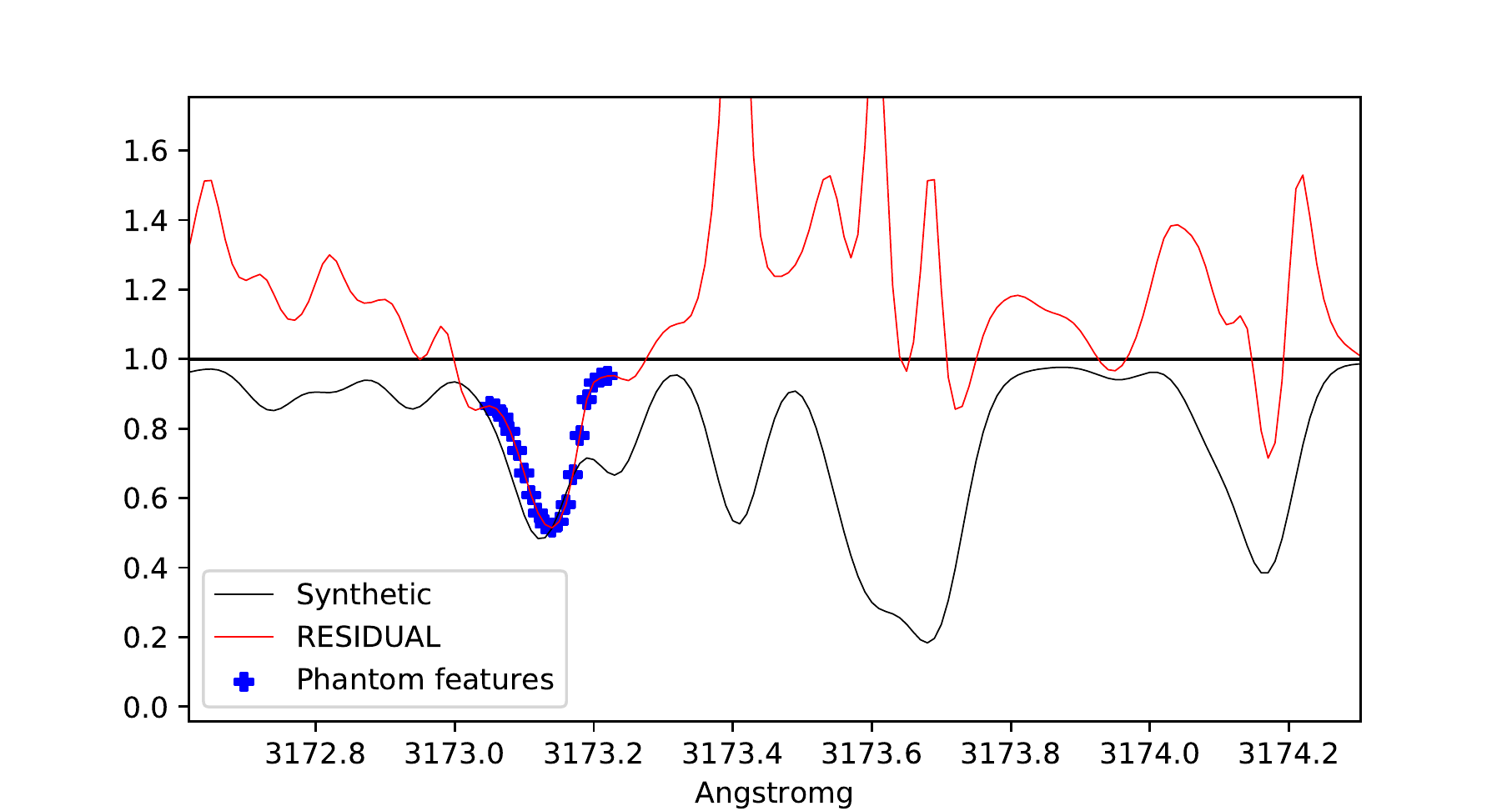}
    \includegraphics[width=8cm]{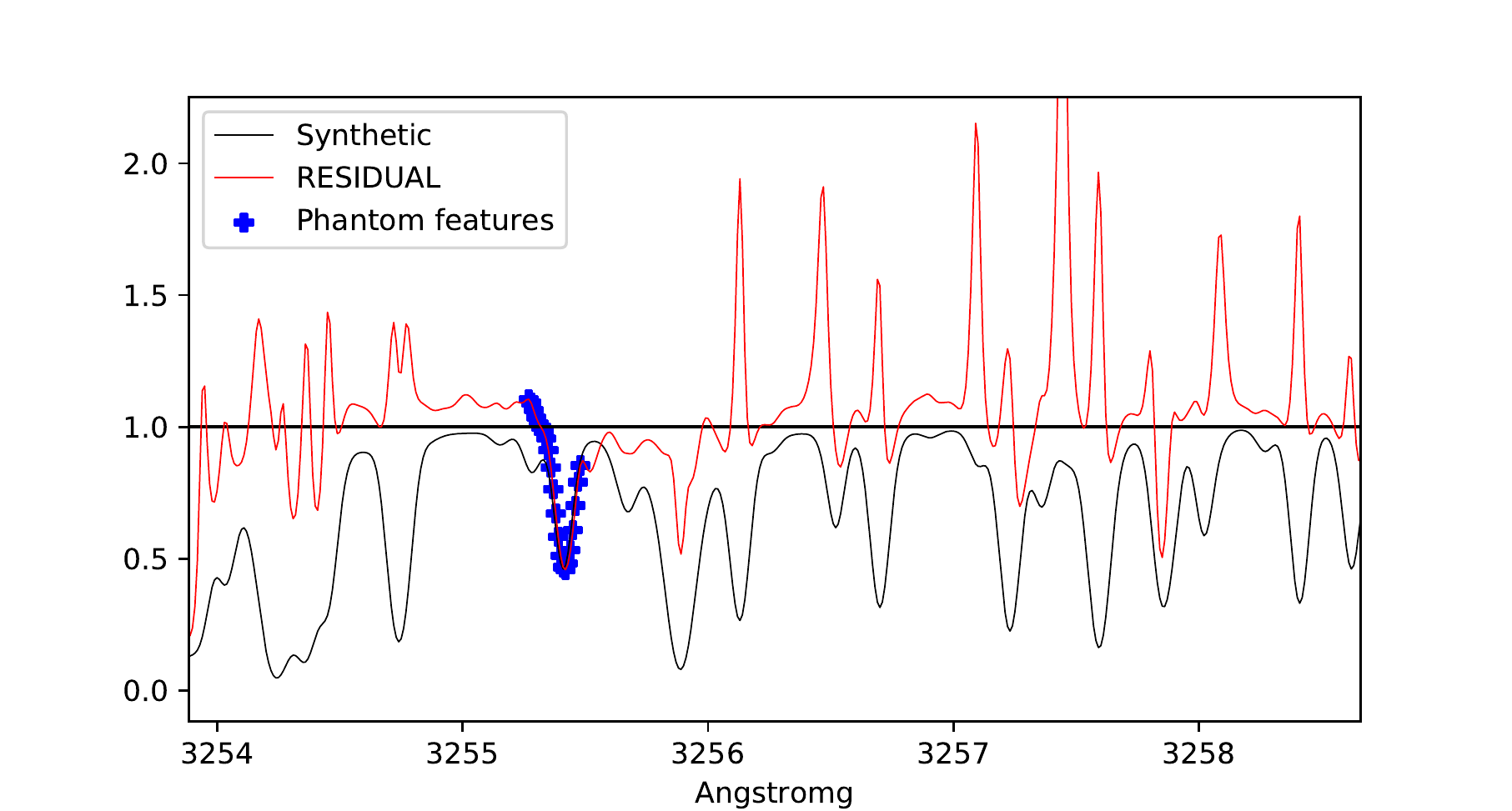}
    \includegraphics[width=8cm]{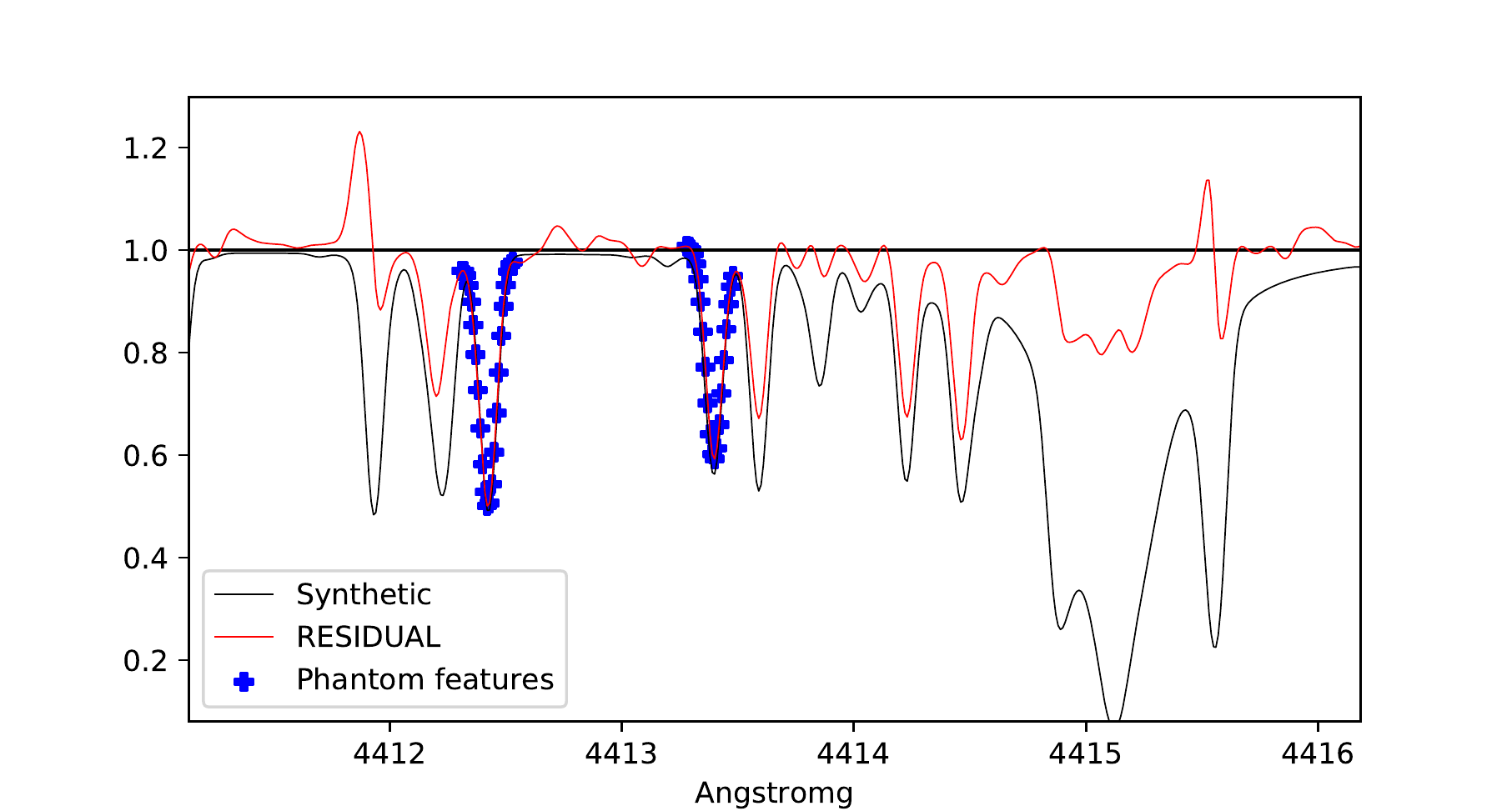}
    \caption{Examples of phantom lines identified with the solar KPNO spectrum. The blue dots layout phantom lines identified in the residual spectrum (red line) by matching it with the synthetic spectrum (black line).}
    \label{fig:phantom}
\end{figure}

\begin{figure*}
    \centering
    \includegraphics[width=17.1cm]{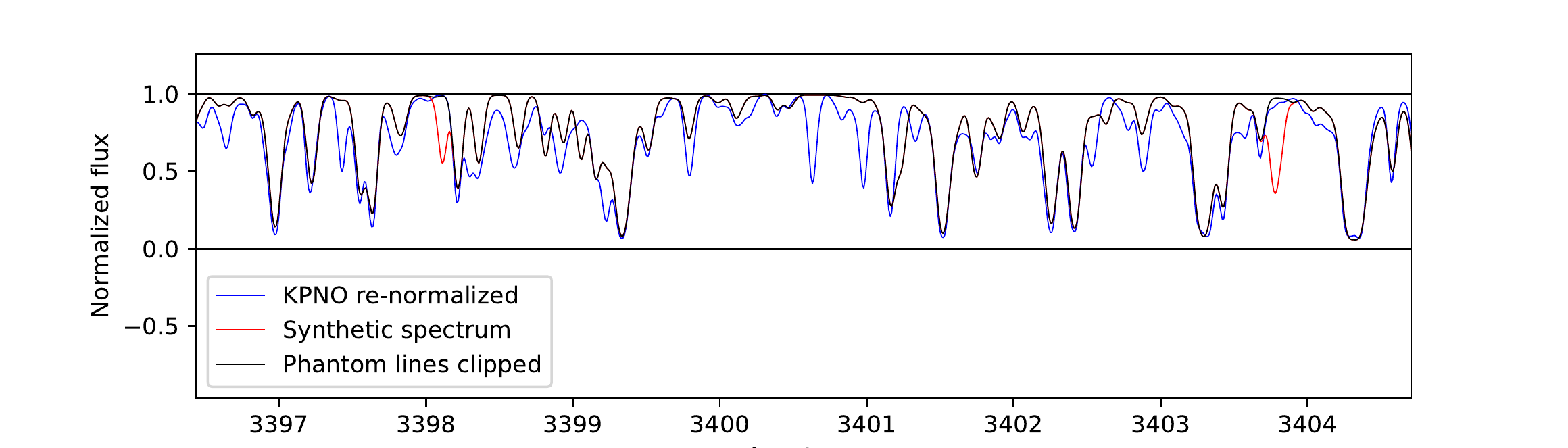}
    \caption{Comparison between the observed (blue), uncorrected synthetic (red), and corrected synthetic (black) solar spectra.}
    \label{fig:NO_PHANTOM}
\end{figure*}
%The synthetic spectrum corrected from phantom lines is represented by the black line.

As the observational template, we used the Kitt Peak National Observatory solar spectrum\footnote{Available at: \url{http://kurucz.harvard.edu/sun/fluxatlas2005/}} (KPNO spectrum henceforth) \cite{kurucz2005MSAIS...8..189K}. Its very high resolution and SNR help to optimize the process of disentangling spectral features. The comparison of the spectra is performed across the wavelength range 2987-9000~\AA; 
the limit in the ultraviolet side set by the KPNO spectrum coverage. This spectrum comes fragmented in several sections, each one with different resolution. 
Namely, we selected flux101 (covering 2987-3351~\AA\ with $R = 348\,000$), flux103a (covering 3232-3952~\AA\ with $R = 348\,480$), flux107 (covering 3806-4950~\AA\ with $R = 340\,500$), flux109 (covering 4299-6116~\AA\ with $R = 521\,360$), flux111 (covering 5643-8271~\AA\ with $R = 522\,900$), and flux113a (covering 7100-9498~\AA\ with $R = 522\,900$). We homogenized their resolution to $R = 340\,500$ and subsequently merged the fragments into a single contiguous spectrum.

Initial comparisons of the KPNO spectrum with the synthetic solar spectrum showed deviations in the continuum level for wavelengths bluer than $\sim$4300~\AA. For the purposes of our analysis, we re-normalized the KPNO merged spectrum by tying it to the synthetic spectrum level in wavelength regions of high flux values without missing or phantom spectral lines. This re-normalized spectrum is made available at Github\footnote{\url{https://github.com/RGiribaldi/FGKstars}}. We remark, however, that the continuum computed by Turbospectrum in these blue wavelengths is likely incorrect. Some known sources of continuum opacity, like NH photodissociation \cite{2015ApJ...809..157F}, have not been taken into account. Moreover, there is some evidence that non-LTE effects are important for computing the blue continuum \cite{2009ApJ...691.1634S,2017ApJ...835..292Y} while the Turbospectrum calculations are performed under LTE.

For the line clipping procedure, we first divided each of the computed synthetic spectra by the KPNO spectrum, producing a residual spectrum that highlights phantom lines as absorption-like features and missing lines as emission-like features. This residual spectrum is then compared with the synthetic spectra of atoms and of each molecular species separately, in a search for matching features. Only lines that are missing in the synthetic spectrum are marked to be removed. Lines with discrepant intensities but that exist in both spectra are kept in the line list. An example of this procedure is depicted in Fig.~\ref{fig:phantom}. A total of 144 features of atomic origin, plus nineteen from C$^{12}$H, five from C$^{12}$N$^{14}$, and fourteen from O$^{16}$H were identified as phantom lines. The other molecular species did not have phantom lines. Figure~\ref{fig:NO_PHANTOM} demonstrates the improvement in the synthetic spectrum obtained with this procedure. Two phantom lines are clearly seen in the red synthetic spectrum at about 3398 and 3404 \AA. It is, however, still a problem that in this wavelength region many lines are missing (blue isolated features). This corrected line list is made available in Github\footnote{\url{https://github.com/RGiribaldi/Master-line-list-for-spectral-synthesis-with-Turbospectrum}}.

\section{Beryllium in turn-off stars of globular clusters}\label{sec:science}

The science case of beryllium abundances in stars of globular clusters was first discussed in \cite{2014Ap&SS.354...55S}, in the context of a previous design study of CUBES \cite{2014Ap&SS.354..191B,2014SPIE.9147E..09B}. As detailed in the Introduction, the goal of this science case is to detect star-to-star variation in the abundance of Be. This will enable studies about the extent of the dilution of nuclear processed material during the process that formed the multiple stellar populations of globular clusters. In the previous study \cite{2014Ap&SS.354...55S}, spectra were simulated for turn-off stars in NGC 6752 and Omega Centauri, with R $\sim$ 20\,000 and four arbitrary SNR levels. Here, we present more realistic simulations of the expected CUBES performance. We make use of the new CUBES simulation tools \cite{genoni_CUBES} to estimate the spectra and SNR expected from actual observations. %of stars in the four clusters listed in the top part of Table \ref{tab:params}.

\subsection{Stellar parameters in the turn-off}

As potential targets, we selected those globular clusters with the brightest turn-offs (TOs) that can be observed from the Paranal observatory (Table \ref{tab:params}). We imposed a limit at $V$ = 18 mag at the TO to select the candidates for observation. When available, photometry and membership information for stars in these clusters were adopted from  \cite{narloch2017MNRAS.471.1446N}.

\begin{table*}
\centering
\normalsize
\caption{Parameters adopted for turn-off stars in nearby clusters.}
\label{tab:params}
\begin{threeparttable}
\footnotesize
\begin{tabular}{l c | c c c c c c c c}
\hline\hline 
Cluster & NGC & Age & [Fe/H] &  $V_{\mathrm{TO}}$ & \teff & \logg & $R_{\mathrm{Sun}}$ & E(B-V) & SNR\\
 & & [Gyr] & [dex] & [mag] & [K] & [dex] & [pc] & [mag] & [1 h.] \\
\hline
NGC 6752 & 6752 & 13.4$^{(a)}$ & $-1.43^{(a)}$ & 17.39$^{(a)}$ & 6120 & 4.20 & 4125$^{(c)}$ & 0.04$^{(a)}$  & 15\\ %& $-1.28^{(d)}$
NGC 6397 & 6397 & 13.5$^{(a)}$ & $-2.03^{(a)}$ & 16.56$^{(a)}$ & 6350 & 4.20 & 2482$^{(c)}$ & 0.18$^{(a)}$  & 25\\ % & $-1.64^{(e)*}$
47 Tuc & 104 & 10.8$^{(a)}$ & $-0.66^{(a)}$ & 17.68$^{(a)}$ & 5940 & 4.25 & 4521$^{(c)}$ & 0.04$^{(a)}$  & 10\\ % & $-0.16^{(f)*}$
%O from -0.91 to -0.06 in (f)
M 4 & 6121 & 10.9 & $-1.20^{(b)}$ & 16.5 & 6177 & 4.19 & 2100 & 0.37 & 25\\ %& $-1.02$ to $-0.69^{(h)}$
\hline
M 12 & 6218 & 12.4 & $-1.48^{(b)}$ & 18.0 & 6216 & 4.18 & 5100 & 0.22 & 10 \\ %& $-1.85$ to $-0.72^{(g)}$
M 55 & 6809 & 13.3 & $-1.81^{(b)}$ & 17.8  & 6290 & 4.18 & 5348 & 0.08 & 10 \\ %& $-1.83$ to $-1.38^{(i)}$
M 22 & 6656 & 12.5 & $-1.64^{(b)}$ & 17.4 & 6253 & 4.18 & 3500 & 0.28 & 10 \\ %& $-1.74$ to $-1.08^{(j)}$
NGC 3201 & 3201 & 10.6 & $-1.35^{(d)}$ & 17.8 & 6200 & 4.20 & 4737 & 0.23 & 10\\ % & $-2.18$ to $-1.34^{(k)}$
M 10 & 6254 & 11.9 & $-1.52^{(b)}$ & 18.0 & 6250 & 4.17 & 5070 & 0.23 & 10 \\ %& $-1.39$ to $-0.91^{(k)}$
%NGC 362 & & & & & & & 8.5 & 0.05\\
$\omega$ Cen & 5139 & 11.0 & $-1.70^{(e)}$ & 18.0 & 6324 & 4.17 & 5426$^{(c)}$ & 0.08 & 10\\
\hline
\end{tabular}
\begin{tablenotes}
\item{} \textbf{Notes.}
Age values determined here likely correspond to the first generation stars only.
$V_{\mathrm{TO}}$ indicates the $V$ magnitude of a turn-off star. $R_{\mathrm{Sun}}$ indicates the distance from the Sun. $E(B-V)$ is the reddening. SNR corresponds to the value measured around the $\lambda3130$~\AA\ line considering 3000s of integration. For the clusters at the bottom part, the variation in magnitude and \teff\ compensate each other, resulting in very similar SNR values.
$^{(a)}$\cite{gratton2003A&A...408..529G},
$^{(b)}$\cite{harris1996AJ....112.1487H},  $^{(c)}$\cite{baumgardt2021MNRAS.505.5957B},
$^{(d)}$\cite{feuillet2021arXiv210512141F},
$^{(e)}$\cite{johnson2020AJ....159..254J}.
\end{tablenotes}
\end{threeparttable}
\end{table*}

\begin{figure*}
    \centering
    \includegraphics[width=5cm]{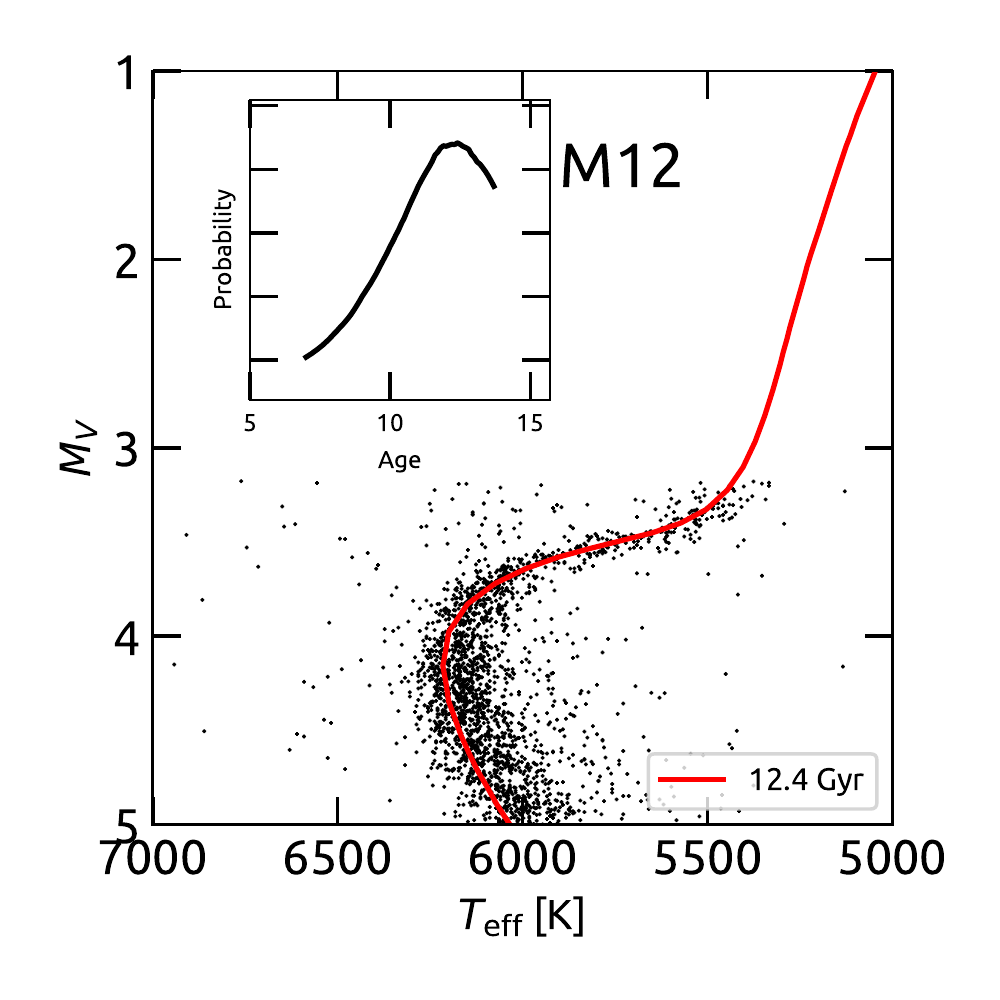}
    \includegraphics[width=5cm]{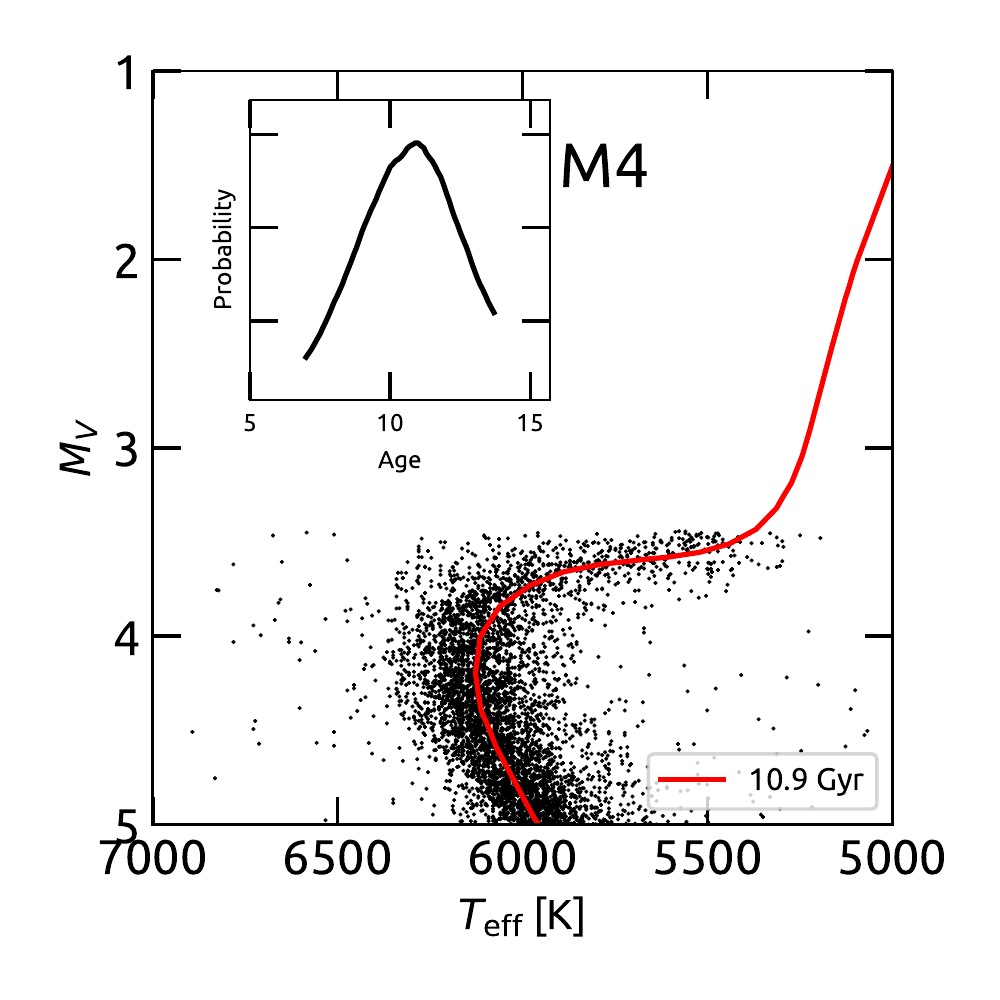}
    \includegraphics[width=5cm]{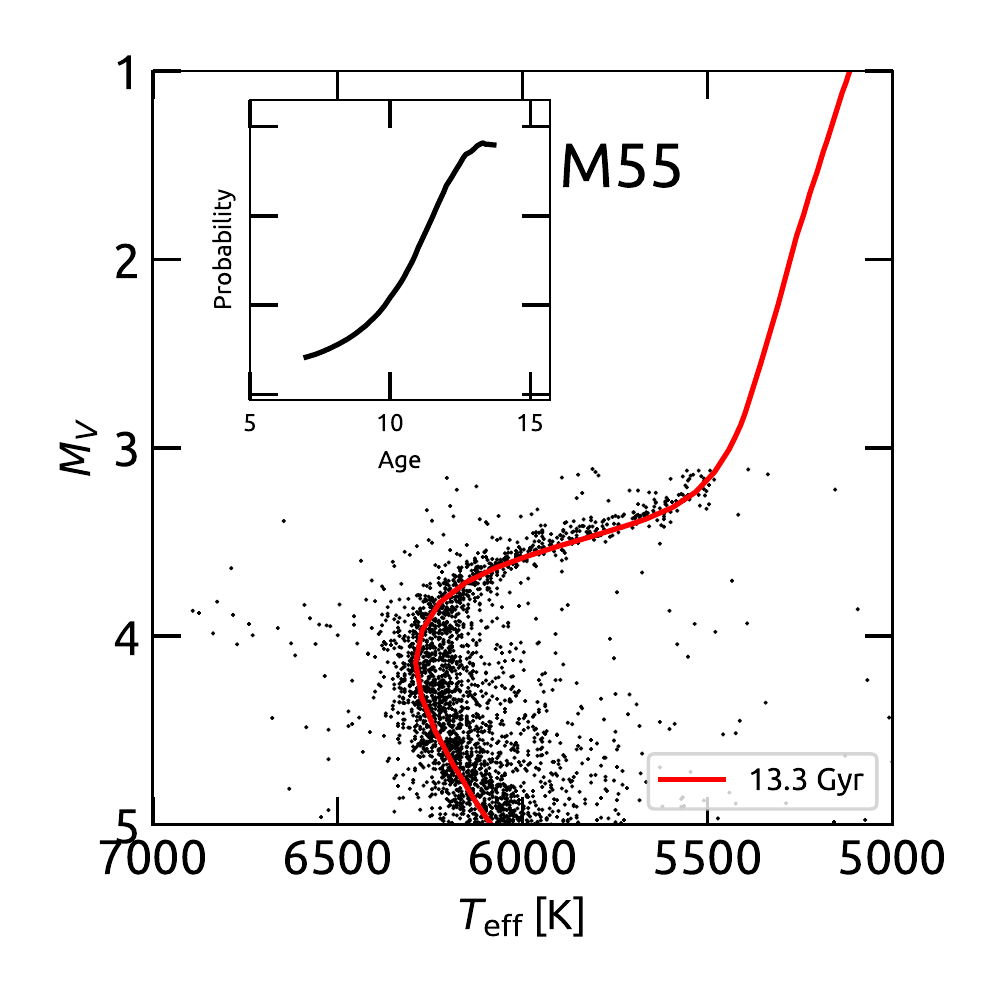}
    \includegraphics[width=5cm]{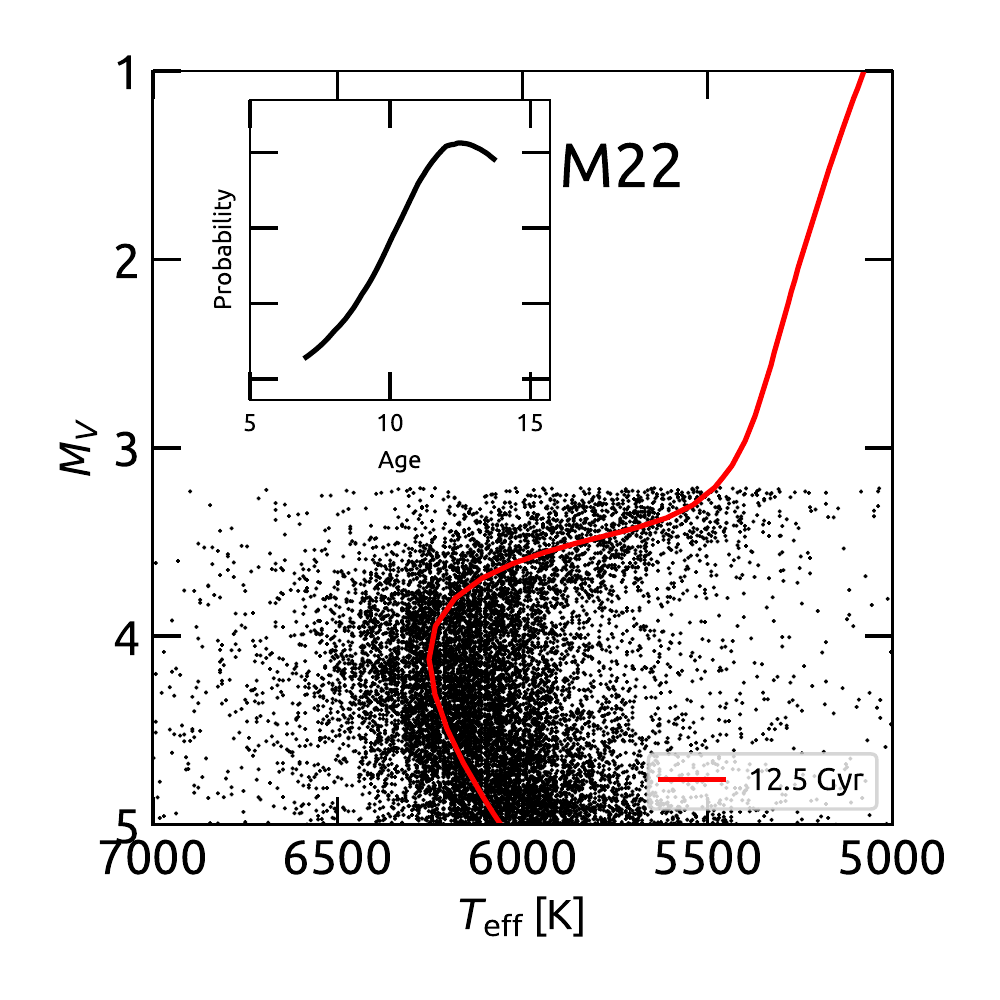}
    \includegraphics[width=5cm]{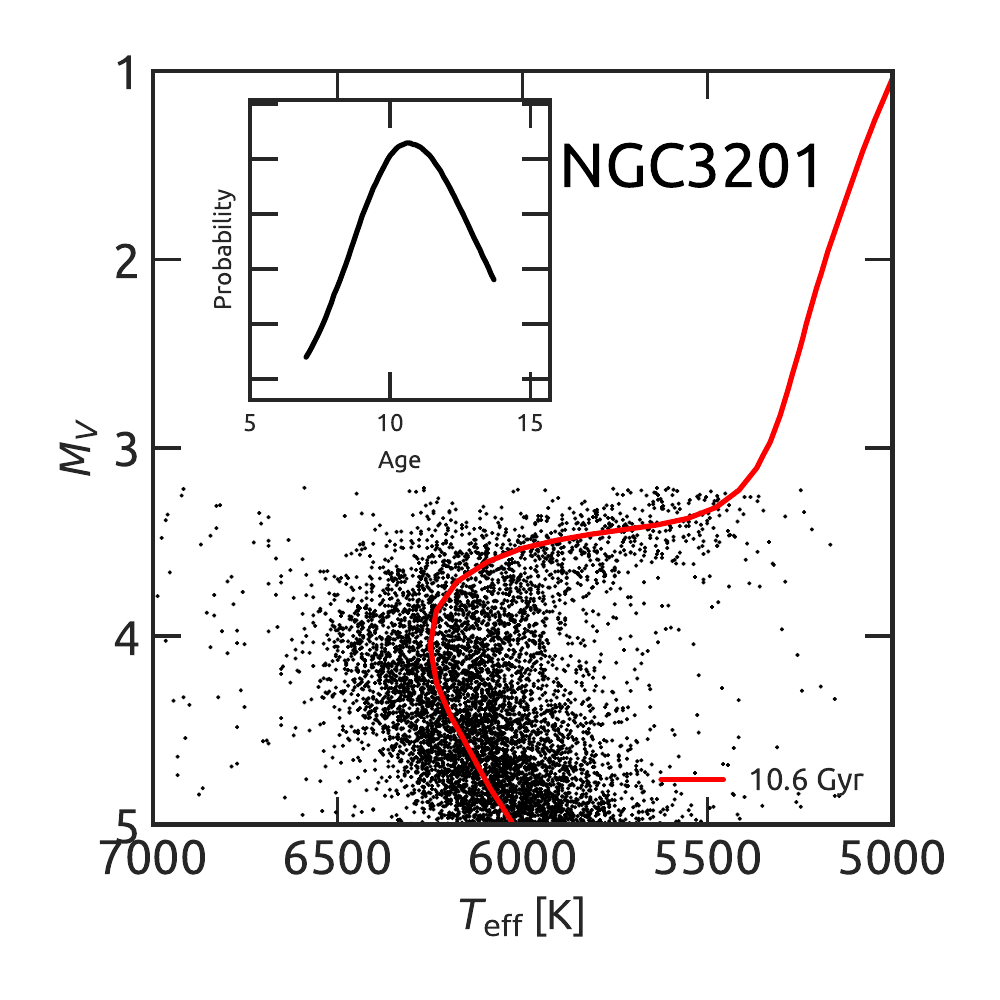}
    \includegraphics[width=5cm]{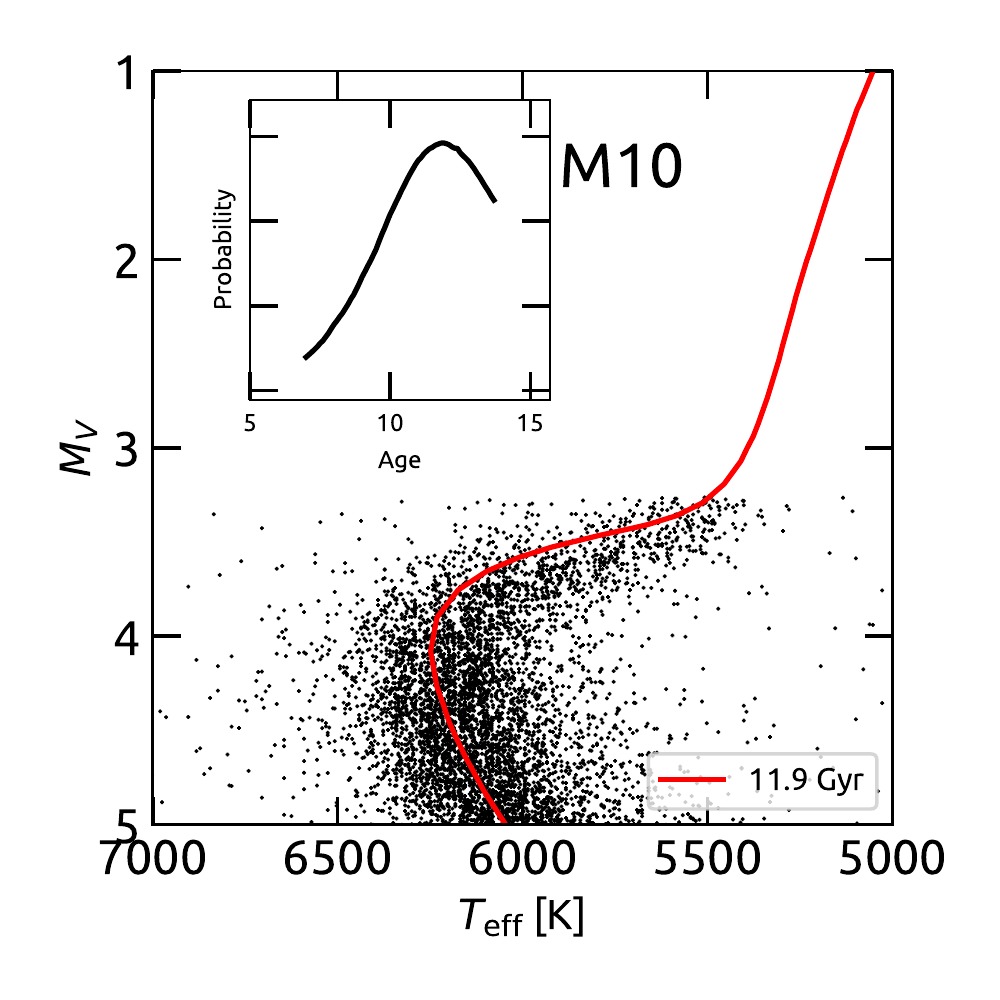}
    \includegraphics[width=5cm]{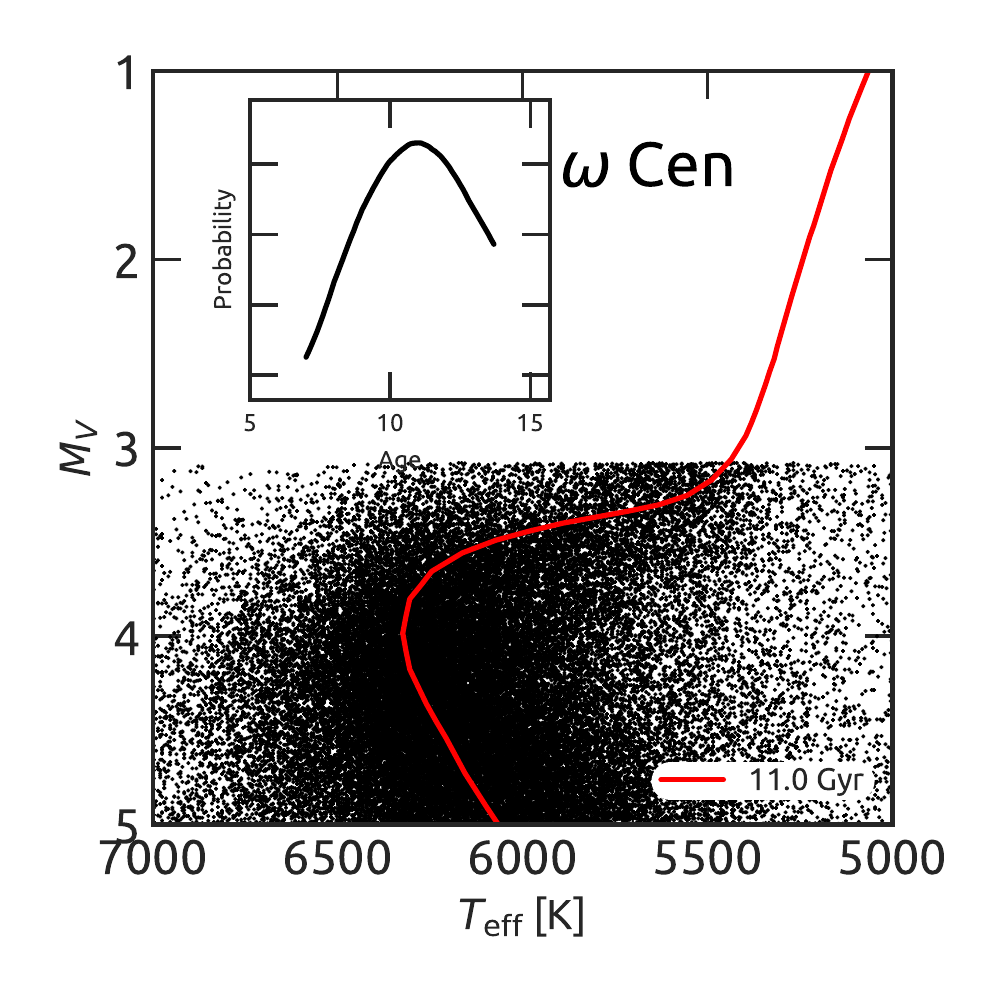}
    \caption{Isochrone fitting to the turn-off of the clusters in the second part of Table \ref{tab:params}.}
    \label{fig:iso_fittings}
\end{figure*}

To simulate the spectra, we need the values of stellar atmospheric parameters and abundances as input. For NGC 6752, NGC 6392, and 47 Tuc, we computed \teff\ and \logg\ of their TO stars using the theoretical isochrones of \cite{yi2003} and the ages and [Fe/H] values listed in Table~\ref{tab:params}, which are compiled along with their sources in the literature. For the remaining clusters, we compiled [Fe/H] and reddening values, $E(B-V)$, from the Harris catalog \cite{harris1996AJ....112.1487H}, which provides relatively well updated information\footnote{https://web.archive.org/web/20110128035351/http://gclusters.altervista.org/}. Although distances are also compiled in that catalog, we decided to use the latest values derived in \cite{baumgardt2021MNRAS.505.5957B}. Effective temperatures were then derived using derredened $(B-V)$ colours and [Fe/H] values with the color--\teff\ calibrations of \cite{casagrande2010A&A...512A..54C}. Ages and \logg\ values were derived by fitting isochones to the distribution of the stars in the Hertzprung-Russel diagram (with $M_V$\footnote{Computed assuming extinction $A_V = 3.1 \times E(B-V)$} vs. \teff). In this process, we allowed distance and $E(B-V)$ to vary as free parameters by at most $\pm5\%$ to improve the fitting. Figure \ref{fig:iso_fittings} shows the isochrones fitted to these clusters. The parameters are all listed in Table \ref{tab:params}. Considering that isochrones with standard He and CNO abundances were used, the ages derived here likely correspond only to stars of the first generation, see \cite{salaris2006ApJ...645.1131S} for a discussion on this topic.

\subsection{Beryllium detection limit with CUBES}

\begin{figure*}
    \centering
    \includegraphics[width=11cm]{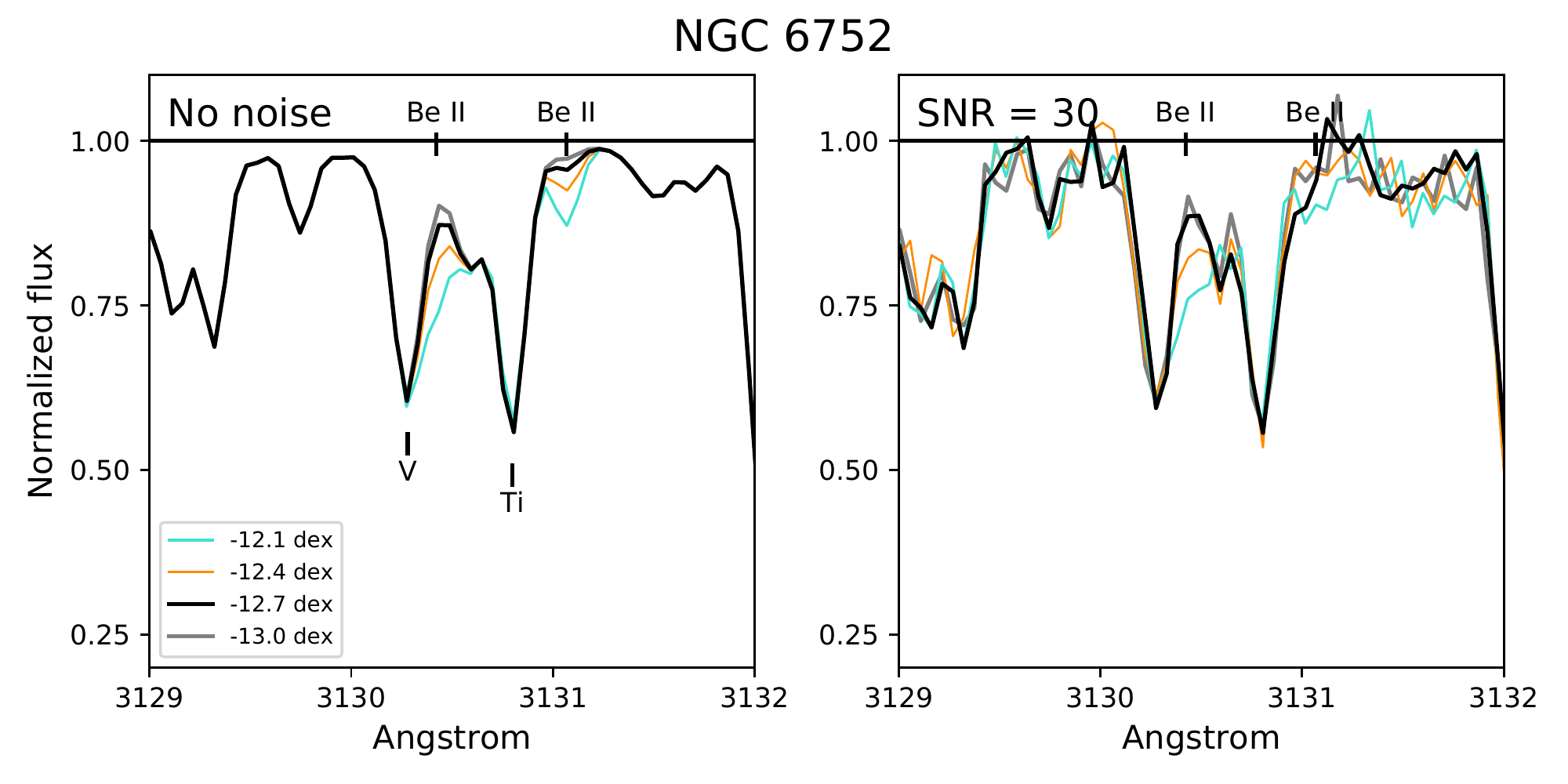}
    \includegraphics[width=11cm]{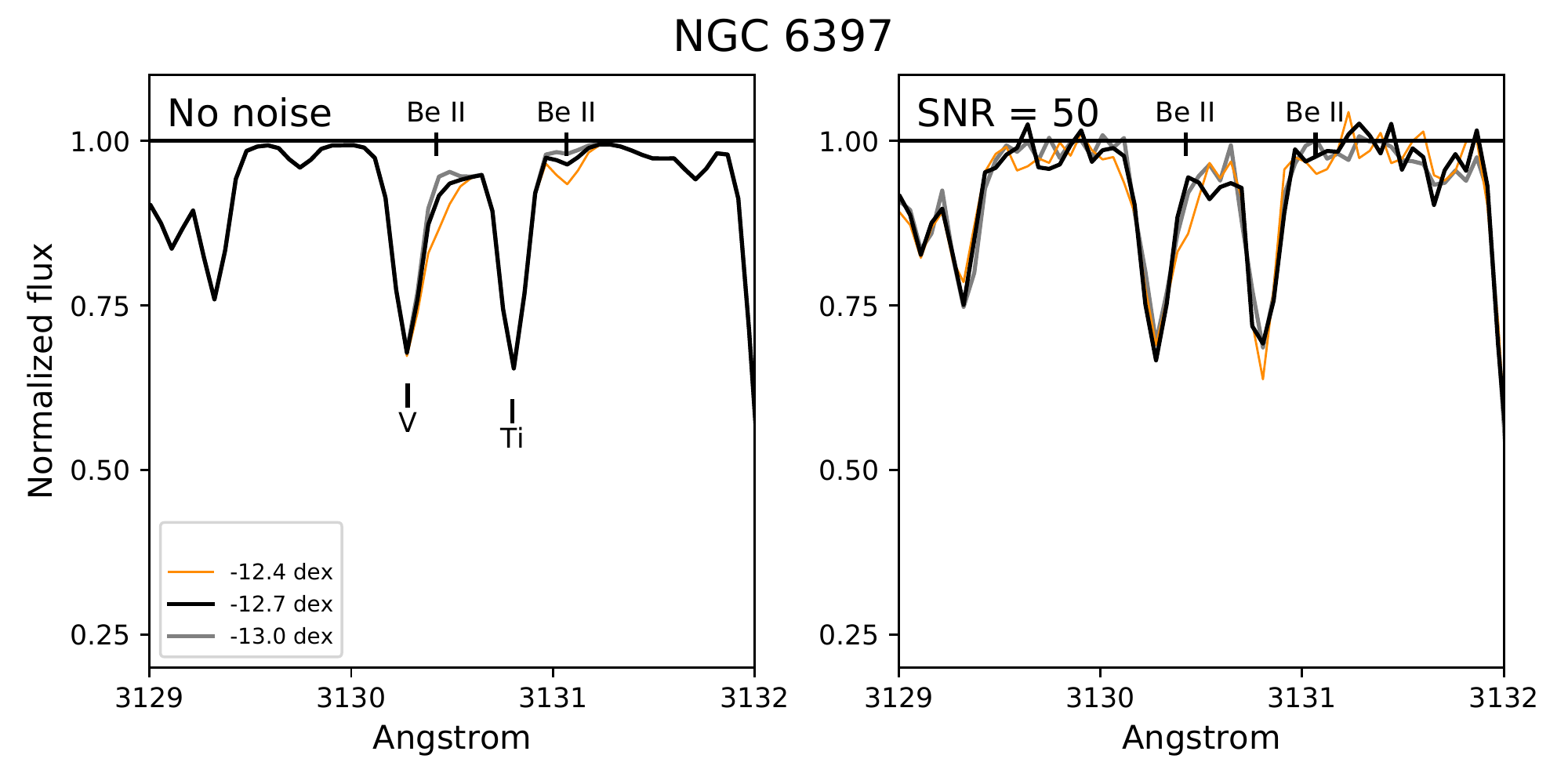}
    \includegraphics[width=11cm]{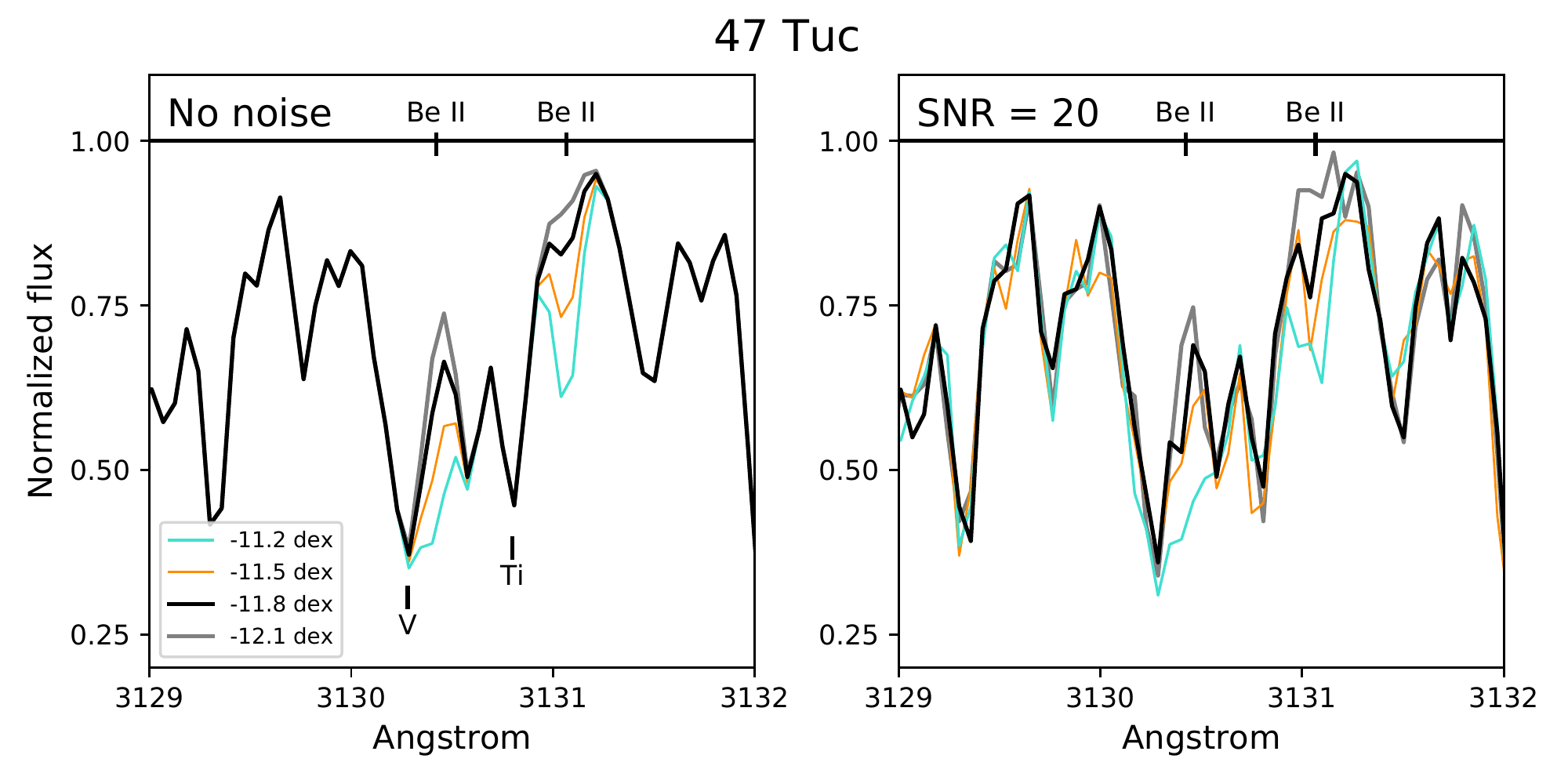}
    \includegraphics[width=11cm]{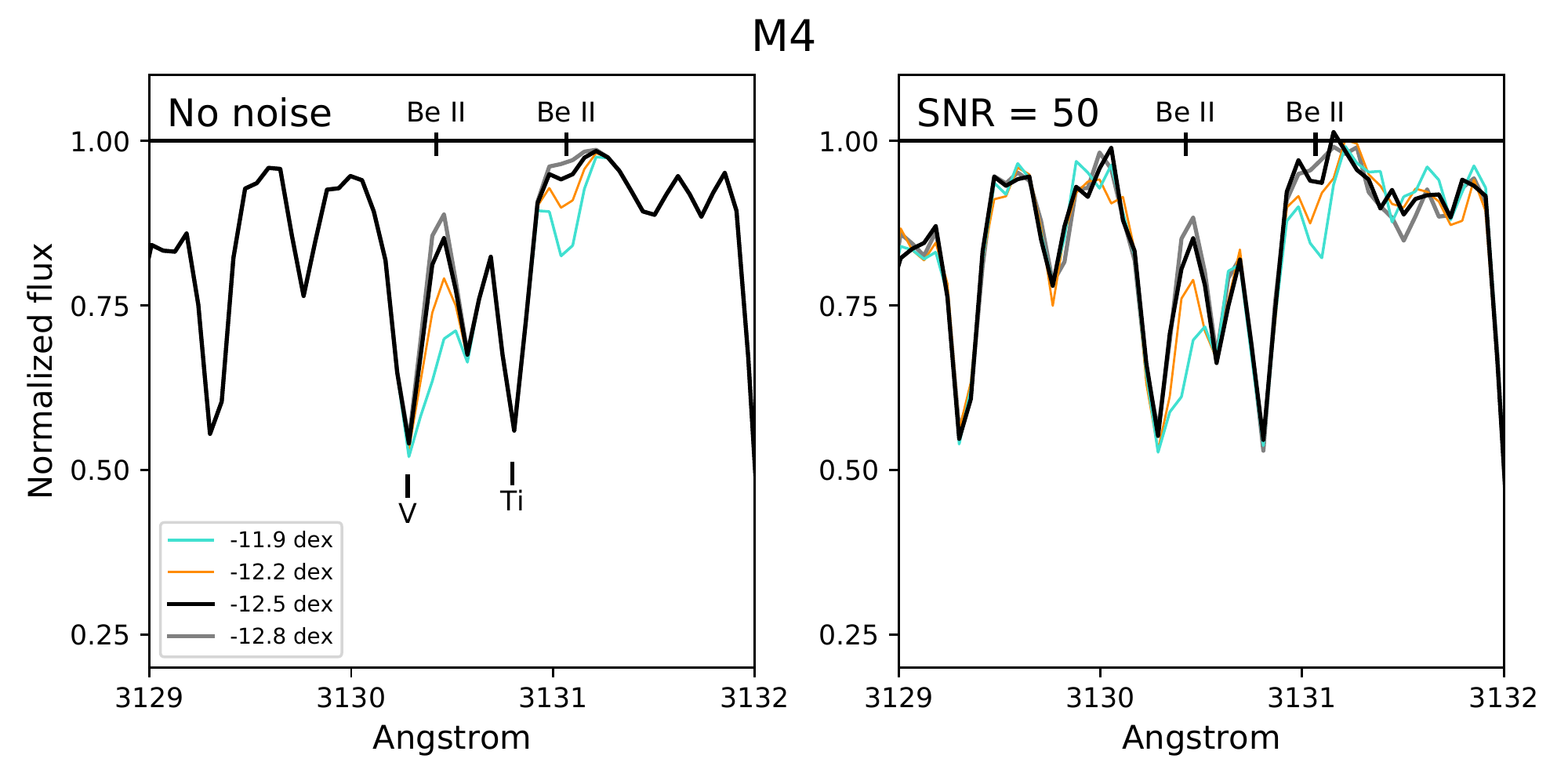}
    \caption{Simulated spectra of turn-off stars in the four globular clusters in the top part of Table~\ref{tab:params}. The panels on the left column show the synthetic spectra without noise. On the right column the same spectra are shown but with noise included at the indicated level. The log(Be/H) abundances are identified as follows: extremely low (gray line), very low (black line), low (orange), and undepleted (cyan).
    }
    \label{fig:Be_features}
\end{figure*}

We discuss the clusters NGC~6752, NGC~6397, 47~Tuc, and M4 as representative cases. The simulated spectra for stars in the turn-off of these clusters are displayed in Fig.~\ref{fig:Be_features}. The panels on the right of the figure show the noised version of the synthetic spectra seen in the left panels. For each cluster, \textbf{three or four} spectra are shown with Be abundances changing in steps of 0.3 dex. In most cases, the undepleted level of Be is considered to be equal to what is observed in the metal-poor field stars of the same metallicity. To estimate this abundance for 47~Tuc, M~4, and NGC 6752 we used the linear relation between Be and [Fe/H] derived in \cite{smiljanic2009A&A...499..103S} and reproduced in Equation\ \ref{eq:be.feh} below.

\begin{equation}
    \log(Be/H) = -10.38\,+\,1.24[Fe/H]\label{eq:be.feh}
\end{equation}

For NGC 6752, this linear relation predicts $\log$(Be/H) = $-$12.1, which is very similar to the detection for one of its turn-off stars, $\log$(Be/H) = $-$12.04, reported in \cite{pasquini2007A&A...464..601P}. For NGC 6397, significantly more metal poor, the linear relation predicts $\log$(Be/H) = -12.9. However, \cite{pasquini2004A&A...426..651P} reports a possible detection at $\log$(Be/H) = -12.35. We thus use this value as the highest possible Be abundance for NGC 6397.

The simulated spectra were computed as described in Section \ref{sec:data}. The SNR values obtained in simulated observation of 3000s are listed in Table~\ref{tab:params}. For the analysis, we assume a final SNR from the stacking of four spectra (which means a total observing time of four hours per star). Thus, the final SNR values, around the Be lines, used in the spectra of Fig.~\ref{fig:Be_features} are 30, 50, 20, and 50 for NGC 6752, NGC 6397, 47~Tuc, and M4, respectively. We recall here that attempts to observe the Be lines in turn-off stars of NGC 6752 and NGC 6397 with UVES required a total 12h to obtain SNR = 8-15 \cite{pasquini2004A&A...426..651P,pasquini2007A&A...464..601P}. The high efficiency of CUBES is demonstrated in this comparison, as it will be possible to obtain twice as much SNR in one third of the time needed with UVES (note, however, the difference in resolution, $\sim$23\,000 with CUBES and $\sim$45\,000 with UVES). Below, we examine each Be line  separately to estimate the minimum abundances possible to be measured from them.

\subsubsection{The redder Be line $\lambda$3131.067~\AA} 

The redder Be line at 3131.067~\AA\ is the weakest of the doublet. However, at the resolution of CUBES, and at the low metallicities of the selected clusters, this line is the cleanest in terms of blends. Figure \ref{fig:Be_features} shows that it can be partially affected by the red wing of the strong nearby line of Ti. 

It seems evident for the older and more metal-poor clusters (NGC 6752 and NGC 6397) that the analysis of this line will be quite challenging. Although the SNR for the case of the NGC 6397 star is almost twice of that for the star in NGC 6752, no spectrum can be clearly distinguished from the others as the Be abundances are all too low for that. Using this line, in a cluster with the metallicity of NGC 6397, only abundances ofs log(Be/H) $>~-12.4$~dex have any chance of being detected. The implication for the proposed science case is that, assuming the undepleted abundance level for these clusters is correct, we can at most detect a dilution of material that reduces the Be abundance by 0.3 dex in clusters with metallicity of [Fe/H] $\sim$ $-$1.4, like NGC 6752. Increasing the total observation time per star from four to 16 hours would double the SNR values that can be obtained. This might improve the detection limit for NGC 6397 (where SNR $\sim$ 100 would be achieved), but would still be of limited effect for NGC 6752 (where SNR $\sim$ 60 would be achieved, not very different from the SNR $\sim$ 50 shown in the case of NGC 6397).

For the younger, more metal-rich clusters, the situation is slightly different. This happens because the Be lines are expected to be stronger, as the undepleted abundance is higher. For 47 Tuc, the spectra are expected to have SNR $\sim$ 20, but it seems to be possible to detect a total variation of up to 0.9 dex in the Be abundances. For the case of M~4, the SNR obtained is higher and the limit of detection seems to be at about log(Be/H) $= -12.5$~dex. This corresponds to a total variation of 0.6 dex in the Be abundance, with respect to the undepleted level.

\subsubsection{The bluer Be line $\lambda$3130.422~\AA} 

The line at 3130.422~\AA\ is the strongest of the doublet but is also seriously blended with a vanadium line, at the resolution of CUBES. Essentially, the Be feature is just a contamination at the wing of the stronger V line. The analysis of this line will also be challenging and will require spectrum synthesis.

For NGC 6752 the spectra with extremely low and very low abundances are indistinguishable. However, it seems clear that the spectra with undepleted abundance can be distinguished from the one with Be lower by 0.3 dex. In addition, the plots also suggest that there is a good chance of bringing the detection limit down to log(Be/H) $\sim -12.7$~dex. This would mean that a total variation of 0.6 dex in Be could potentially be detected. For NGC 6397, it seems that no clear detection can be made.

For the two clusters of higher metallicity, 47~Tuc and M~4, it seems more clear that a variation of 0.6 dex from the undepleted level can be detected. A larger variation might be detectable in 47 Tuc, in particular if the SNR is improved with longer total exposures.

\section{Summary}\label{sec:summary}

In the context of the phase A study of CUBES, we investigated the possibility of detecting variations of the surface Be abundance in turn off stars of globular clusters. Beryllium abundances have the potential of providing new key insight into the mixture of ejected and pristine material that happened during the formation of the multiple stellar generations inside globular clusters.

The preliminary design of CUBES \cite{Zanutta_CUBES}, indicates that the spectra around the Be lines will have R $\sim$ 23\,000 and a sampling of 2.35 pixels. We estimated the SNR that can be obtained using the CUBES E2E \cite{genoni_CUBES} and assuming observing blocks with exposure times of 3000s. To simulate the observations, we used synthetic spectra computed with a line list that has been cleaned from spurious lines. This line list is made publicly available, together with a re-normalized version of the solar spectrum \cite{kurucz2005MSAIS...8..189K}, which has been homogenized in resolution covering the region between 2987 to 9498 \AA. 

We discuss simulated observations of stars in four clusters (NGC 6752, NGC 6593, 47 Tuc, and M4), but present an extended list of 10 nearby clusters that can be considered as potential targets for CUBES observations. These clusters have limiting turn-off magnitude of $V$ = 18 mag and are accessible from the Paranal observatory. The analysis indicates that, in clusters with [Fe/H] $\gtrsim$ $-$1.4, a total decrease of about 0.6 dex from the undepleted Be abundance level can be detected. The analysis becomes more challenging for clusters of lower metallicity where the Be lines are weaker. For NGC 6397, with [Fe/H] = -2.0, there seems little chance of detecting any scatter in Be. In cases of higher metallicity clusters, with stronger Be lines, if better SNR can be obtained then it might be possible to detect a variation of up to 0.9 dex. Such analyses, however, will always be challenging. At the resolution of CUBES, the lines are never free of blends. Nevertheless, this is a science case that can only be explored with CUBES. Other current near-UV spectrographs would require prohibitive large exposures to achieve similar results.

\begin{acknowledgements}
Use was made of the Simbad database, operated at the CDS, Strasbourg, France, and of NASA’s Astrophysics Data System Bibliographic Services.
This work has made use of the VALD database, operated at Uppsala University, the Institute of Astronomy RAS in Moscow, and the University of Vienna. 
\end{acknowledgements}

\section*{\small Data availability}
\small
The datasets generated during the current study are available in the GitHub repository. A merged, re-normalised, re-sampled, and resolution homogenised version of the Kitt Peak Solar Atlas: \url{https://github.com/RGiribaldi/FGKstars}.
A corrected line list with format compatible with Turbospectrum:\\ \url{https://github.com/RGiribaldi/Master-line-list-for-spectral-synthesis-with-Turbospectrum}.

\section*{\small Conflict of Interest}
The authors declare that they have no conflict of interest.
This work was supported by the National Science Centre, Poland, through project 2018/31/B/ST9/01469.


\bibliographystyle{spphys}       % APS-like style for physics
\bibliography{apssamp}   % name your BibTeX data base

\providecommand{\noopsort}[1]{}\providecommand{\singleletter}[1]{#1}%
\begin{thebibliography}{10}
\providecommand{\url}[1]{{#1}}
\providecommand{\urlprefix}{URL }
\expandafter\ifx\csname urlstyle\endcsname\relax
  \providecommand{\doi}[1]{DOI \discretionary{}{}{}#1}\else
  \providecommand{\doi}{DOI \discretionary{}{}{}\begingroup
  \urlstyle{rm}\Url}\fi

\bibitem{1978ApJ...223..487C}
J.G. {Cohen}, \apj \textbf{223}, 487 (1978).
\newblock \doi{10.1086/156284}

\bibitem{2018ARA&A..56...83B}
N.~{Bastian}, C.~{Lardo}, \araa \textbf{56}, 83 (2018).
\newblock \doi{10.1146/annurev-astro-081817-051839}

\bibitem{1999Natur.402...55L}
Y.W. {Lee}, J.M. {Joo}, Y.J. {Sohn}, S.C. {Rey}, H.C. {Lee}, A.R. {Walker},
  \nat \textbf{402}(6757), 55 (1999).
\newblock \doi{10.1038/46985}

\bibitem{2020A&ARv..28....5C}
S.~{Cassisi}, M.~{Salaris}, \aapr \textbf{28}(1), 5 (2020).
\newblock \doi{10.1007/s00159-020-00127-y}

\bibitem{bedin2004ApJ...605L.125B}
L.R. {Bedin}, G.~{Piotto}, J.~{Anderson}, S.~{Cassisi}, I.R. {King},
  Y.~{Momany}, G.~{Carraro}, \apj \textbf{605}(2), L125 (2004).
\newblock \doi{10.1086/420847}

\bibitem{piotto2007ApJ...661L..53P}
G.~{Piotto}, L.R. {Bedin}, J.~{Anderson}, I.R. {King}, S.~{Cassisi}, A.P.
  {Milone}, S.~{Villanova}, A.~{Pietrinferni}, A.~{Renzini}, \apj
  \textbf{661}(1), L53 (2007).
\newblock \doi{10.1086/518503}

\bibitem{kraft1994PASP..106..553K}
R.P. {Kraft}, PASP \textbf{106}, 553 (1994).
\newblock \doi{10.1086/133416}

\bibitem{gratton2007A&A...464..953G}
R.G. {Gratton}, S.~{Lucatello}, A.~{Bragaglia}, E.~{Carretta}, S.~{Cassisi},
  Y.~{Momany}, E.~{Pancino}, E.~{Valenti}, V.~{Caloi}, R.~{Claudi},
  F.~{D'Antona}, S.~{Desidera}, P.~{Fran{\c{c}}ois}, G.~{James}, S.~{Moehler},
  S.~{Ortolani}, L.~{Pasquini}, G.~{Piotto}, A.~{Recio-Blanco}, \aap
  \textbf{464}(3), 953 (2007).
\newblock \doi{10.1051/0004-6361:20066061}

\bibitem{carretta2015ApJ...810..148C}
E.~{Carretta}, \apj \textbf{810}(2), 148 (2015).
\newblock \doi{10.1088/0004-637X/810/2/148}

\bibitem{pancino2017A&A...601A.112P}
E.~{Pancino}, D.~{Romano}, B.~{Tang}, G.~{Tautvai{\v{s}}ien{\.{e}}}, A.R.
  {Casey}, P.~{Gruyters}, D.~{Geisler}, I.~{San Roman}, S.~{Randich}, E.J.
  {Alfaro}, A.~{Bragaglia}, E.~{Flaccomio}, A.J. {Korn}, A.~{Recio-Blanco},
  R.~{Smiljanic}, G.~{Carraro}, A.~{Bayo}, M.T. {Costado}, F.~{Damiani},
  P.~{Jofr{\'e}}, C.~{Lardo}, P.~{de Laverny}, L.~{Monaco}, L.~{Morbidelli},
  L.~{Sbordone}, S.G. {Sousa}, S.~{Villanova}, \aap \textbf{601}, A112 (2017).
\newblock \doi{10.1051/0004-6361/201730474}

\bibitem{2012ApJ...760...86C}
J.G. {Cohen}, E.N. {Kirby}, \apj \textbf{760}(1), 86 (2012).
\newblock \doi{10.1088/0004-637X/760/1/86}

\bibitem{2020MNRAS.492.1641M}
S.~{M{\'e}sz{\'a}ros}, T.~{Masseron}, D.A. {Garc{\'\i}a-Hern{\'a}ndez},
  C.~{Allende Prieto}, T.C. {Beers}, D.~{Bizyaev}, D.~{Chojnowski}, R.E.
  {Cohen}, K.~{Cunha}, F.~{Dell'Agli}, G.~{Ebelke}, J.G.
  {Fern{\'a}ndez-Trincado}, P.~{Frinchaboy}, D.~{Geisler}, S.~{Hasselquist},
  F.~{Hearty}, J.~{Holtzman}, J.~{Johnson}, R.R. {Lane}, I.~{Lacerna},
  P.~{Longa-Pe{\~n}a}, S.R. {Majewski}, S.L. {Martell}, D.~{Minniti},
  D.~{Nataf}, D.L. {Nidever}, K.~{Pan}, R.P. {Schiavon}, M.~{Shetrone}, V.V.
  {Smith}, J.S. {Sobeck}, G.S. {Stringfellow}, L.~{Szigeti}, B.~{Tang}, J.C.
  {Wilson}, O.~{Zamora}, \mnras \textbf{492}(2), 1641 (2020).
\newblock \doi{10.1093/mnras/stz3496}

\bibitem{2021A&A...646A...9C}
E.~{Carretta}, A.~{Bragaglia}, \aap \textbf{646}, A9 (2021).
\newblock \doi{10.1051/0004-6361/202039392}

\bibitem{1981ApJ...245L..79C}
P.L. {Cottrell}, G.S. {Da Costa}, \apjl \textbf{245}, L79 (1981).
\newblock \doi{10.1086/183527}

\bibitem{2009A&A...499..835V}
P.~{Ventura}, F.~{D'Antona}, \aap \textbf{499}(3), 835 (2009).
\newblock \doi{10.1051/0004-6361/200811139}

\bibitem{2007A&A...464.1029D}
T.~{Decressin}, G.~{Meynet}, C.~{Charbonnel}, N.~{Prantzos}, S.~{Ekstr{\"o}m},
  \aap \textbf{464}(3), 1029 (2007).
\newblock \doi{10.1051/0004-6361:20066013}

\bibitem{2013A&A...552A.121K}
M.~{Krause}, C.~{Charbonnel}, T.~{Decressin}, G.~{Meynet}, N.~{Prantzos}, \aap
  \textbf{552}, A121 (2013).
\newblock \doi{10.1051/0004-6361/201220694}

\bibitem{2009A&A...507L...1D}
S.E. {de Mink}, O.R. {Pols}, N.~{Langer}, R.G. {Izzard}, \aap \textbf{507}(1),
  L1 (2009).
\newblock \doi{10.1051/0004-6361/200913205}

\bibitem{2019A&ARv..27....8G}
R.~{Gratton}, A.~{Bragaglia}, E.~{Carretta}, V.~{D'Orazi}, S.~{Lucatello},
  A.~{Sollima}, \aapr \textbf{27}(1), 8 (2019).
\newblock \doi{10.1007/s00159-019-0119-3}

\bibitem{2002A&A...390...91B}
P.~{Bonifacio}, L.~{Pasquini}, F.~{Spite}, A.~{Bragaglia}, E.~{Carretta},
  V.~{Castellani}, M.~{Centuri{\`o}n}, A.~{Chieffi}, R.~{Claudi},
  G.~{Clementini}, F.~{D'Antona}, S.~{Desidera}, P.~{Fran{\c{c}}ois}, R.G.
  {Gratton}, F.~{Grundahl}, G.~{James}, S.~{Lucatello}, C.~{Sneden},
  O.~{Straniero}, \aap \textbf{390}, 91 (2002).
\newblock \doi{10.1051/0004-6361:20020620}

\bibitem{2010A&A...524L...2S}
Z.X. {Shen}, P.~{Bonifacio}, L.~{Pasquini}, S.~{Zaggia}, \aap \textbf{524}, L2
  (2010).
\newblock \doi{10.1051/0004-6361/201015738}

\bibitem{2010MNRAS.402L..72V}
P.~{Ventura}, F.~{D'Antona}, \mnras \textbf{402}(1), L72 (2010).
\newblock \doi{10.1111/j.1745-3933.2010.00805.x}

\bibitem{2014ApJ...791...39D}
V.~{D'Orazi}, G.C. {Angelou}, R.G. {Gratton}, J.C. {Lattanzio}, A.~{Bragaglia},
  E.~{Carretta}, S.~{Lucatello}, Y.~{Momany}, \apj \textbf{791}(1), 39 (2014).
\newblock \doi{10.1088/0004-637X/791/1/39}

\bibitem{2012A&A...542A..67P}
N.~{Prantzos}, \aap \textbf{542}, A67 (2012).
\newblock \doi{10.1051/0004-6361/201219043}

\bibitem{pasquini2004A&A...426..651P}
L.~{Pasquini}, P.~{Bonifacio}, S.~{Randich}, D.~{Galli}, R.G. {Gratton}, \aap
  \textbf{426}, 651 (2004).
\newblock \doi{10.1051/0004-6361:20041254}

\bibitem{pasquini2007A&A...464..601P}
L.~{Pasquini}, P.~{Bonifacio}, S.~{Randich}, D.~{Galli}, R.G. {Gratton},
  B.~{Wolff}, \aap \textbf{464}(2), 601 (2007).
\newblock \doi{10.1051/0004-6361:20066260}

\bibitem{2014A&A...563A...3P}
L.~{Pasquini}, A.~{Koch}, R.~{Smiljanic}, P.~{Bonifacio}, A.~{Modigliani}, \aap
  \textbf{563}, A3 (2014).
\newblock \doi{10.1051/0004-6361/201323220}

\bibitem{2000SPIE.4008..534D}
H.~{Dekker}, S.~{D'Odorico}, A.~{Kaufer}, B.~{Delabre}, H.~{Kotzlowski}, in
  \emph{Optical and IR Telescope Instrumentation and Detectors}, \emph{Society
  of Photo-Optical Instrumentation Engineers (SPIE) Conference Series}, vol.
  4008, ed. by M.~{Iye}, A.F. {Moorwood} (2000), \emph{Society of Photo-Optical
  Instrumentation Engineers (SPIE) Conference Series}, vol. 4008, pp. 534--545.
\newblock \doi{10.1117/12.395512}

\bibitem{Zanutta_CUBES}
A.~{Zanutta}, D.~{Atkinson}, V.~{Baldini}, {et al.}, ExA  (2021)

\bibitem{Calcines_CUBES}
A.~{Calcines}, M.~{Wells}, K.~{O\'Brien}, {et al.}, ExA  (2021)

\bibitem{2014Ap&SS.354..121P}
L.~{Pasquini}, \apss \textbf{354}(1), 121 (2014).
\newblock \doi{10.1007/s10509-014-2049-x}

\bibitem{2016SPIE.9908E..9JE}
C.J. {Evans}, M.~{Puech}, M.~{Rodrigues}, B.~{Barbuy}, J.G. {Cuby},
  G.~{Dalton}, E.~{Fitzsimons}, F.~{Hammer}, P.~{Jagourel}, L.~{Kaper}, S.L.
  {Morris}, T.J. {Morris}, in \emph{Ground-based and Airborne Instrumentation
  for Astronomy VI}, \emph{Society of Photo-Optical Instrumentation Engineers
  (SPIE) Conference Series}, vol. 9908, ed. by C.J. {Evans}, L.~{Simard},
  H.~{Takami} (2016), \emph{Society of Photo-Optical Instrumentation Engineers
  (SPIE) Conference Series}, vol. 9908, p. 99089J.
\newblock \doi{10.1117/12.2231675}

\bibitem{2018SPIE10702E..2EE}
C.J. {Evans}, B.~{Barbuy}, B.~{Castilho}, R.~{Smiljanic}, J.~{Melendez},
  J.~{Japelj}, S.~{Cristiani}, C.~{Snodgrass}, P.~{Bonifacio}, M.~{Puech},
  A.~{Quirrenbach}, in \emph{Ground-based and Airborne Instrumentation for
  Astronomy VII}, \emph{Society of Photo-Optical Instrumentation Engineers
  (SPIE) Conference Series}, vol. 10702, ed. by C.J. {Evans}, L.~{Simard},
  H.~{Takami} (2018), \emph{Society of Photo-Optical Instrumentation Engineers
  (SPIE) Conference Series}, vol. 10702, p. 107022E.
\newblock \doi{10.1117/12.2312022}

\bibitem{2020SPIE11447E..60E}
H.~{Ernandes}, C.J. {Evans}, B.~{Barbuy}, B.~{Castilho}, G.~{Cescutti},
  N.~{Christlieb}, S.~{Cristiani}, P.~{Di Marcantonio}, C.~{Hansen},
  A.~{Quirrenbach}, R.~{Smiljanic}, in \emph{Society of Photo-Optical
  Instrumentation Engineers (SPIE) Conference Series}, \emph{Society of
  Photo-Optical Instrumentation Engineers (SPIE) Conference Series}, vol. 11447
  (2020), \emph{Society of Photo-Optical Instrumentation Engineers (SPIE)
  Conference Series}, vol. 11447, p. 1144760.
\newblock \doi{10.1117/12.2562497}

\bibitem{turbospectrum}
B.~{Plez}.
\newblock {Turbospectrum: Code for spectral synthesis} (2012)

\bibitem{gustafson2008}
B.~{Gustafsson}, B.~{Edvardsson}, K.~{Eriksson}, U.G. {J{\o}rgensen},
  {\AA}.~{Nordlund}, B.~{Plez}, A\&A \textbf{486}, 951 (2008).
\newblock \doi{10.1051/0004-6361:200809724}

\bibitem{Ryabchikova2015PhyS...90e4005R}
T.~{Ryabchikova}, N.~{Piskunov}, R.L. {Kurucz}, H.C. {Stempels}, U.~{Heiter},
  Y.~{Pakhomov}, P.S. {Barklem}, Physica Scripta \textbf{90}(5), 054005 (2015).
\newblock \doi{10.1088/0031-8949/90/5/054005}

\bibitem{kurucz1992RMxAA..23...45K}
R.L. {Kurucz}, Revista Mexicana de Astronomia y Astrofisica \textbf{23}, 45
  (1992)

\bibitem{masseron2014A&A...571A..47M}
T.~{Masseron}, B.~{Plez}, S.~{Van Eck}, R.~{Colin}, I.~{Daoutidis},
  M.~{Godefroid}, P.F. {Coheur}, P.~{Bernath}, A.~{Jorissen}, N.~{Christlieb},
  \aap \textbf{571}, A47 (2014).
\newblock \doi{10.1051/0004-6361/201423956}

\bibitem{brooke2014ApJS..210...23B}
J.S.A. {Brooke}, R.S. {Ram}, C.M. {Western}, G.~{Li}, D.W. {Schwenke}, P.F.
  {Bernath}, \apjs \textbf{210}(2), 23 (2014).
\newblock \doi{10.1088/0067-0049/210/2/23}

\bibitem{sneden2014ApJS..214...26S}
C.~{Sneden}, S.~{Lucatello}, R.S. {Ram}, J.S.A. {Brooke}, P.~{Bernath}, \apjs
  \textbf{214}(2), 26 (2014).
\newblock \doi{10.1088/0067-0049/214/2/26}

\bibitem{brooke2013JQSRT.124...11B}
J.S.A. {Brooke}, P.F. {Bernath}, T.W. {Schmidt}, G.B. {Bacskay}, \jqsrt
  \textbf{124}, 11 (2013).
\newblock \doi{10.1016/j.jqsrt.2013.02.025}

\bibitem{peterson2020A&A...638A..64P}
R.C. {Peterson}, B.~{Barbuy}, M.~{Spite}, \aap \textbf{638}, A64 (2020).
\newblock \doi{10.1051/0004-6361/202037689}

\bibitem{genoni_CUBES}
M.~{Genoni}, M.~{Landoni}, G.~{Cupani}, {et al.}, ExA  (2021)

\bibitem{bell1994MNRAS.268..771B}
R.A. {Bell}, G.~{Paltoglou}, M.J. {Tripicco}, \mnras \textbf{268}, 771 (1994).
\newblock \doi{10.1093/mnras/268.3.771}

\bibitem{jofre2014A&A...564A.133J}
P.~{Jofr{\'e}}, U.~{Heiter}, C.~{Soubiran}, S.~{Blanco-Cuaresma}, C.C.
  {Worley}, E.~{Pancino}, T.~{Cantat-Gaudin}, L.~{Magrini}, M.~{Bergemann},
  J.I. {Gonz{\'a}lez Hern{\'a}ndez}, V.~{Hill}, C.~{Lardo}, P.~{de Laverny},
  K.~{Lind}, T.~{Masseron}, D.~{Montes}, A.~{Mucciarelli}, T.~{Nordlander},
  A.~{Recio Blanco}, J.~{Sobeck}, R.~{Sordo}, S.G. {Sousa}, H.~{Tabernero},
  A.~{Vallenari}, S.~{Van Eck}, \aap \textbf{564}, A133 (2014).
\newblock \doi{10.1051/0004-6361/201322440}

\bibitem{kurucz2005MSAIS...8..189K}
R.L. {Kurucz}, Memorie della Societa Astronomica Italiana Supplementi
  \textbf{8}, 189 (2005)

\bibitem{2015ApJ...809..157F}
J.M. {Fontenla}, P.C. {Stancil}, E.~{Landi}, \apj \textbf{809}(2), 157 (2015).
\newblock \doi{10.1088/0004-637X/809/2/157}

\bibitem{2009ApJ...691.1634S}
C.I. {Short}, P.H. {Hauschildt}, \apj \textbf{691}(2), 1634 (2009).
\newblock \doi{10.1088/0004-637X/691/2/1634}

\bibitem{2017ApJ...835..292Y}
M.E. {Young}, C.I. {Short}, \apj \textbf{835}(2), 292 (2017).
\newblock \doi{10.3847/1538-4357/835/2/292}

\bibitem{2014Ap&SS.354...55S}
R.~{Smiljanic}, \apss \textbf{354}(1), 55 (2014).
\newblock \doi{10.1007/s10509-014-1916-9}

\bibitem{2014Ap&SS.354..191B}
B.~{Barbuy}, V.~{Bawden Macanhan}, P.~{Bristow}, B.~{Castilho}, H.~{Dekker},
  B.~{Delabre}, M.~{Diaz}, C.~{Gneiding}, F.~{Kerber}, H.~{Kuntschner}, G.~{La
  Mura}, W.~{Maciel}, J.~{Mel{\'e}ndez}, L.~{Pasquini}, C.B. {Pereira},
  P.~{Petitjean}, R.~{Reiss}, C.~{Siqueira-Mello}, R.~{Smiljanic}, J.~{Vernet},
  \apss \textbf{354}(1), 191 (2014).
\newblock \doi{10.1007/s10509-014-2039-z}

\bibitem{2014SPIE.9147E..09B}
P.~{Bristow}, B.~{Barbuy}, V.B. {Macanhan}, B.~{Castilho}, H.~{Dekker},
  B.~{Delabre}, M.~{Diaz}, C.~{Gneiding}, F.~{Kerber}, H.~{Kuntschner}, G.~{La
  Mura}, R.~{Reiss}, J.~{Vernet}, in \emph{Ground-based and Airborne
  Instrumentation for Astronomy V}, \emph{Society of Photo-Optical
  Instrumentation Engineers (SPIE) Conference Series}, vol. 9147, ed. by S.K.
  {Ramsay}, I.S. {McLean}, H.~{Takami} (2014), \emph{Society of Photo-Optical
  Instrumentation Engineers (SPIE) Conference Series}, vol. 9147, p. 914709.
\newblock \doi{10.1117/12.2054751}

\bibitem{narloch2017MNRAS.471.1446N}
W.~{Narloch}, J.~{Kaluzny}, R.~{Poleski}, M.~{Rozyczka}, W.~{Pych}, I.B.
  {Thompson}, \mnras \textbf{471}(2), 1446 (2017).
\newblock \doi{10.1093/mnras/stx1637}

\bibitem{gratton2003A&A...408..529G}
R.G. {Gratton}, A.~{Bragaglia}, E.~{Carretta}, G.~{Clementini}, S.~{Desidera},
  F.~{Grundahl}, S.~{Lucatello}, \aap \textbf{408}, 529 (2003).
\newblock \doi{10.1051/0004-6361:20031003}

\bibitem{harris1996AJ....112.1487H}
W.E. {Harris}, Astrophysical Journal \textbf{112}, 1487 (1996).
\newblock \doi{10.1086/118116}

\bibitem{baumgardt2021MNRAS.505.5957B}
H.~{Baumgardt}, E.~{Vasiliev}, \mnras \textbf{505}(4), 5957 (2021).
\newblock \doi{10.1093/mnras/stab1474}

\bibitem{feuillet2021arXiv210512141F}
D.K. {Feuillet}, C.L. {Sahlholdt}, S.~{Feltzing}, L.~{Casagrande}, arXiv
  e-prints arXiv:2105.12141 (2021)

\bibitem{johnson2020AJ....159..254J}
C.I. {Johnson}, A.K. {Dupree}, M.~{Mateo}, I.~{Bailey}, John~I., E.W.
  {Olszewski}, M.G. {Walker}, The Astronomical Journal \textbf{159}(6), 254
  (2020).
\newblock \doi{10.3847/1538-3881/ab8819}

\bibitem{yi2003}
S.K. {Yi}, Y.C. {Kim}, P.~{Demarque}, ApJS \textbf{144}, 259 (2003).
\newblock \doi{10.1086/345101}

\bibitem{casagrande2010A&A...512A..54C}
L.~{Casagrande}, I.~{Ram{\'\i}rez}, J.~{Mel{\'e}ndez}, M.~{Bessell},
  M.~{Asplund}, \aap \textbf{512}, A54 (2010).
\newblock \doi{10.1051/0004-6361/200913204}

\bibitem{salaris2006ApJ...645.1131S}
M.~{Salaris}, A.~{Weiss}, J.W. {Ferguson}, D.J. {Fusilier}, \apj
  \textbf{645}(2), 1131 (2006).
\newblock \doi{10.1086/504520}

\bibitem{smiljanic2009A&A...499..103S}
R.~{Smiljanic}, L.~{Pasquini}, P.~{Bonifacio}, D.~{Galli}, R.G. {Gratton},
  S.~{Randich}, B.~{Wolff}, \aap \textbf{499}(1), 103 (2009).
\newblock \doi{10.1051/0004-6361/200810592}

\end{thebibliography}

%\begin{thebibliography}{}

%\end{thebibliography}

\end{document}